%% file: root.tex
\documentclass[letterpaper, 10pt, conference]{ieeeconf}      

\IEEEoverridecommandlockouts                              

\input{packages.tex}

\usepackage{eso-pic}
\AddToShipoutPictureFG*{
  \AtPageLowerLeft{
    \put(72,35){
      \parbox{450pt}{
        \centering
        \scriptsize
        This work has been submitted to the IEEE for possible publication.
        Copyright may be transferred without notice, after which this
        version may no longer be accessible.
      }
    }
  }
}

\title{\LARGE \bf
Automated Design and Evaluation Pipeline for Optimal Control Transcriptions in Model Predictive Control
}

\author{Fabio Schneider$^{1}$, Peter Kötting$^{1}$, Renato Loureiro$^{2}$
\thanks{$^{1}$German Aerospace Center, Institute of Robotics and Mechatronics, Münchener Str. 20, 82234 Wessling, Germany.}%
\thanks{$^{2}$University of Stuttgart, Institute of Flight Mechanics and Controls, Pfaffenwaldring 27, 70569 Stuttgart, Germany.}
}

\begin{document}

\maketitle
\thispagestyle{empty}
\pagestyle{empty}

\begin{abstract}
The computational burden of Model Predictive Control represents a bottleneck limiting its widespread deployment in real-time applications. The selection of the optimal control transcription strategy for a given MPC formulation is a key driver of computational efficiency. Identifying performant MPC transcriptions requires significant effort and domain knowledge. 
We propose an automated pipeline for the design and evaluation of performant optimal control transcriptions for a given MPC formulation. Custom multi-objective design metrics, such as computation time or closed-loop performance, can be specified and are minimized by the pipeline. As a transcription method, a pseudospectral Legendre-Gauss-Radau method is employed. We show that it incorporates forward Euler integration and implicit collocation methods of low and high order in one formulation, leveraging the pipeline to evaluate both numerical approaches. Furthermore, different mesh configurations can be parameterized. To identify the optimal configuration of the transcription, a Bayesian Optimization framework is applied. The pipeline’s improvements in MPC performance are demonstrated through On-Orbit Servicing scenarios involving spacecraft motion planning and closed-loop control.
\end{abstract}

\section{INTRODUCTION}
Model Predictive Control (MPC) is an established control methodology that repeatedly solves an optimal control problem (OCP) to determine the next control input. One of the limiting factors for the broader application of MPC is the computational cost~\cite{li_model_2026}. The numerical scheme and the OCP formulation are key components in determining the computational complexity~\cite{rawlings_model_2017,kelly_introduction_2017}, leading to the question of which OCP transcription best matches the specific performance metrics. 

Direct methods are the default approach to solve the OCP in the context of MPC~\cite{grune_nonlinear_2017}. A numerical discretization scheme is used to transcribe the continuous OCP into a Nonlinear Program (NLP)~\cite{betts_survey_1998}. Two prominent direct methods are Multiple-Shooting (MS)~\cite{bock_multiple_1984,betts_survey_1998} and collocation methods~\cite{rao_survey_2010}.

MS is a time-marching formulation that solves an initial value problem (IVP). The time domain is divided into several sub-intervals, and the IVP is solved simultaneously on each of them. The continuity of the solution trajectory is ensured by continuity constraints at the interval borders. The order of the integration scheme determines the solution accuracy. A common class of integration methods is explicit and implicit Runge-Kutta methods. In the MPC context, MS transcription is often used with explicit integration schemes, such as in automotive~\cite{duijkeren_RealTimeNMPCSemiAutomated_2015}, aerospace~\cite{reiter_TemporalCascadingPlanning_2026} and robotics~\cite{nam_RealTimeControlConstrainedDDP_2026}.

Collocation methods enforce that the system dynamics match the derivative of an approximate state trajectory at given collocation points, resulting in an implicit formulation~\cite{rao_survey_2010}. Thus, a system of equations must be solved in each iteration. A common approach is to use pseudospectral methods~\cite{kameswaran_ConvergenceRatesDirect_2008,garg_direct_2011,garg_unified_2010,darby_direct_2011}. The time domain is discretized using non-equidistant collocation points derived from the roots of an orthogonal polynomial~\cite{garg_direct_2011}. Pseudospectral methods are often used for offline trajectory optimization. In recent years, pseudospectral transcriptions are employed more frequently in MPC formulations, e.g., in space applications~\cite{sun_guidance_2025}, or in the automotive field~\cite{gao_accurate_2023,allamaa_safety_2023}.
\begin{figure}[htbp]
        \centering
        \input{figures/schematic_methodology/schematic_methodology.tex}
        \caption{Automated Design and Evaluation Pipeline for Optimal Control Transcriptions (APOCT)\label{fig:bo-pipeline}}
\end{figure}
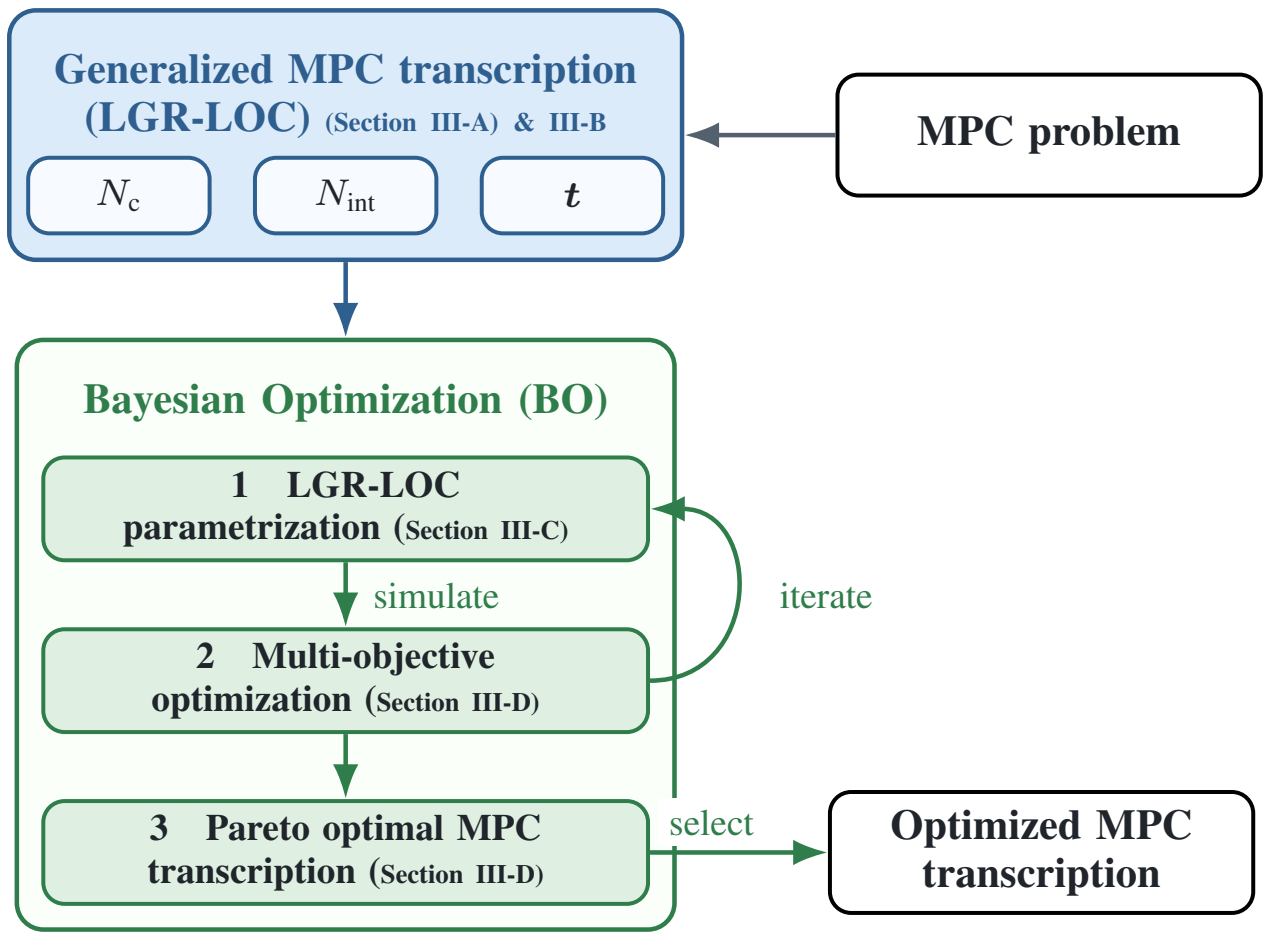

In application-driven MPC design, key decisions include choosing between explicit and implicit formulations, selecting the approximation order, and determining the mesh structure. Several surveys analyze and compare optimal control transcriptions, discussing their respective advantages and disadvantages in the context of MPC~\cite{rawlings_model_2017,sanchez_mpc_2017,diehl_fast_2006,rao_survey_2010}. However, general guidelines for selecting a suitable OCP formulation for custom MPC problems remain difficult to establish. This is primarily due to the strong interdependence between the system dynamics, the embedded constraints, the solver, and the accuracy requirements.
Benchmarking these approaches is time-consuming, as it demands full implementation and scenario-based comparison for each parametrization. The question arises whether a general approach exists for finding the best-performing optimal control transcription for a given MPC scenario?

\newpage
An automated design and evaluation pipeline for optimal control transcriptions (APOCT) is proposed to identify an optimized formulation for a given MPC problem, evaluated against custom performance metrics. This is enabled by the integration of a local pseudospectral Legendre-Gauss-Radau (LGR-LOC)~\cite{garg_direct_2011,darby_direct_2011} method, which is proven to reduce, in a first-order parametrization, to an explicit multiple-shooting formulation with forward Euler integration. Furthermore, the transcription enables a variable local mesh structure by allowing the number of sub-intervals, the interval boundaries, and the prediction horizon to vary. This provides a high degree of flexibility in discretizing the OCP. The LGR-LOC is embedded in a Bayesian Optimization pipeline, shown in Fig.~\ref{fig:bo-pipeline}. To the best of the authors' knowledge, the integration of a flexible OCP transcription into a hyperparameter optimization pipeline for identifying performant MPC formulations is a novel contribution.

The proposed APOCT is demonstrated on On-Orbit Servicing (OOS) examples. We execute detumbling scenarios similar to those in~\cite{kotting_esagnc_2026}. The pipeline is applied to examples with motion planning MPC (open-loop) and closed-loop MPC.

The remainder of this paper is organized as follows. The continuous OCP and the LGR-LOC transcription are introduced in Section~\ref{sec:background}. Section~\ref{sec:APOCT} presents the APOCT, derives the explicit and implicit formulations contained in the LGR-LOC, and details the Bayesian optimization of the transcription parameters. The resulting pipeline is evaluated on the OOS examples in Section~\ref{sec:results}.

The notation is introduced in the following. Vectors are denoted by bold $\bm v$ and matrices by bold non-italic $\mathbf{M}$. Let $\mathbb{R}$ denote the set of real numbers and $\mathbb{I} := \{1,2,3,\ldots\}$ the set of positive integers. For $a \leq b$, we define $\mathbb{R}_{[a,b]} := [a,b]$ and $\mathbb{I}_{[a,b]} := \mathbb{I} \cap [a,b]$. We denote variables $(\cdot)_i^l$, where the subscript $i\in \mathbb{I}_{[1,N_\text{c}]}$ denotes the collocation point and the superscript $l\in \mathbb{I}_{[1,N_\text{int}]}$ denotes the interval. $N_\text{c}$ specifies the number of collocation nodes per interval and $N_\text{int}$ the total number of sub-intervals. The start time of interval $l$ is denoted by $t^l$. The terminal state is denoted by $(\cdot)^{N_\text{int}+1}$.

\section{Background}\label{sec:background}
In this section, we describe the continuous OCP and present an NLP based on the LGR-LOC method, which is embedded in the APOCT.
\subsection{Optimal control}
A general continuous-time optimal control problem is given by 
\begin{subequations}\label{eq:continous-ocp}
  \begin{align}
  \min_{\bm u(\cdot)} \quad & V_\text{f}(\bm x(T)) + \int\limits_0^T  \mathcal{L}(\bm x(t), \bm  u(t), t) \, \mathrm dt \label{eq:ocp_J}\\
  \text{subject to} \quad & \dot{\bm x}(t) = \bm f(\bm x(t),\bm u(t), t), \quad t \in \mathbb{R}_{[0,T]} \label{eq:ocp_x_dot}\\
  & \bm x(0) = \bm x_\text{start}, \label{eq:ocp_x_start}\\
  & \bm g(\bm x(t), \bm u(t), t) \leq 0, \quad t \in \mathbb{R}_{[0,T]} \label{eq:ocp_g}\\
  & \bm x(T) \in \mathcal{X}_\text{f} \label{eq:ocp_x_terminal_region}
   \end{align}
\end{subequations}
with the (nonlinear) system dynamics~\eqref{eq:ocp_x_dot}, the initial condition~\eqref{eq:ocp_x_start}, general state and control input constraints~\eqref{eq:ocp_g}, and the terminal constraint~\eqref{eq:ocp_x_terminal_region} enforcing that the terminal state $\bm x(T)$ lies within the terminal region $\mathcal{X}_\text{f}$. The cost functional~\eqref{eq:ocp_J} consists of the terminal cost term $V_\text{f}(\bm x(T))$, which penalizes deviation from the desired final state, and the stage cost term expressed as the integral of the Lagrangian $\mathcal{L}(\bm x(t), \bm u(t), t)$ over the time horizon $T$.

\subsection{Local Legendre-Gauss-Radau (LGR-LOC) method}\label{subsec:background}
We use a sparsity-exploiting formulation of Legendre-Gauss-Radau (LGR) collocation on multiple domains, referred to as LGR-LOC to emphasize the local, non-global scheme~\cite{kameswaran_ConvergenceRatesDirect_2008,garg_direct_2011,garg_unified_2010,darby_direct_2011}. It enables exploiting the sparsity structure with the Fatrop solver~\cite{vanroye_fatrop_2023}. We assume constant control inputs within each interval. This yields a control discretization that is coarser than or equal to the MPC sampling period~\cite{bayat_lgrmpc_2023}.

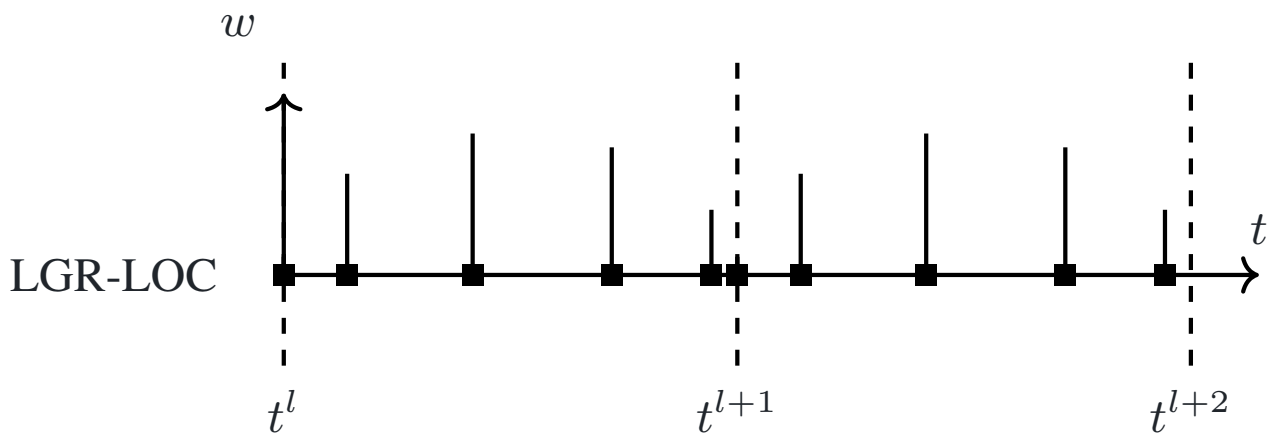
\begin{figure}[htb]
  \input{figures/lgrloc_nodes_schematic/ps_lgl_lgr_plot_loc.tex}
  \caption{Roots of Legendre Polynomial including the left endpoint shown for two local intervals $N_\text{int} =2$.\label{fig:lgrloc_nodes}}
\end{figure}

The LGR-LOC method uses Gauss-Radau quadrature to discretize the continuous-time cost functional~\eqref{eq:ocp_J}. Legendre polynomials of degree $N_\text{c}$ are used to derive the nodes for numerical integration~\cite{wang_symplectic_2021}
\begin{align}
  P_{N_\text{c}}(\tau) = \frac{1}{2^{N_\text{c}} {N_\text{c}}!} \frac{\mathrm d^{N_\text{c}}}{\mathrm d \tau^{N_\text{c}}}  \left[\left(\tau^2 - 1\right)^{{N_\text{c}}}\right].
\end{align}
When using the Legendre-Gauss-Radau integration scheme, the roots of the polynomial $P_{N_\text{c}} + P_{{N_\text{c}}-1}$, including the left endpoint of the interval, are used. The resulting $N_\text{c}$ non-equidistant roots $\tau \in [-1,1)$ are shown in Fig.~\ref{fig:lgrloc_nodes} for two intervals. The roots $\tau_i$ are transformed to an arbitrary interval $[t^l,t^{l+1}]$ via
\begin{align}
  \tau_i \mapsto t_{i}^l = \frac{t^{l+1} - t^l}{2} \tau_i + \frac{t^{l+1}+t^{l}}{2}.\label{eq:t-tau-trafo}
\end{align}

\begin{figure}
  \includegraphics{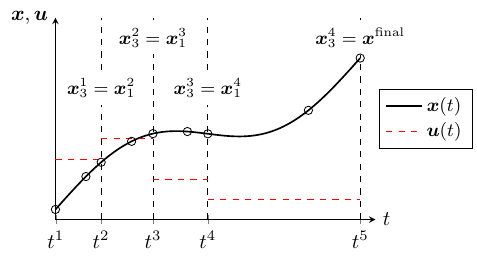}
  \caption{Schematic overview for state and control input discretization for $N_\text{c} = 2$ and $N_\text{int} = 4$.\label{fig:xuapproximation}}
\end{figure}

The interval borders for each interval are grouped in the vector $\bm t = (t^1,\ldots, t^{N_\text{int}+1})^\text{T}$. 
In addition to the $N_\text{c}$ collocation points $\tau_1,\ldots,\tau_{N_\text{c}}$, there is another node $\tau_{{N_\text{c}}+1}$ that is not a collocation point but is used to interpolate the terminal state. The state is approximated in each interval using Lagrange polynomials as basis functions
\begin{align}
  \bm x^l(\tau)  = \sum\limits_{i=1}^{{N_\text{c}}+1} \bm x^l_{i} L_{i}(\tau)\label{eq:lgr-x-u}
\end{align}
with
\begin{align*}
  L_{i}(\tau) = \prod_{j=1,\, j\neq i}^{{N_\text{c}}+1} \frac{\tau-\tau_{j}}{\tau_{i}-\tau_{j}}, \qquad \forall i=1,\ldots, {N_\text{c}}+1.
\end{align*}
The control input is constant within each interval, $\bm u^l(\tau) = \bm{u}^l$. Fig.~\ref{fig:xuapproximation} shows the control and state approximation over several intervals.

The dynamic constraints are evaluated at each non-equidistant collocation point $\tau_j$ and, using transformation~\eqref{eq:t-tau-trafo}, are given by
\begin{align}
\dot{\bm x}^l(\tau_j) &=  \frac{\mathrm{d}}{\mathrm{d}t} \left(\sum\limits_{i=1}^{{N_\text{c}}+1} \bm x_{i}^l L_{i}^{{N_\text{c}}+1}(\tau_j)\right)\notag \\ &= \frac{2}{t^{l+1}-t^l}\sum\limits_{i=1}^{{N_\text{c}}+1} \bm x_{i}^l \dot L_{i}^{{N_\text{c}}+1}(\tau_j) \overset{!}{=} \bm f(\bm x_j^l,\bm u^l, t^l_j)\label{eq:theory-ps-lgl-colloc-constraint}
\end{align} 
with the system dynamics evaluated at time $t_j^l$. Equation~\eqref{eq:theory-ps-lgl-colloc-constraint}, evaluated for $\tau_1,\ldots,\tau_{N_\text{c}}$, is written in matrix form as 
\begin{align}
  \frac{2}{t^{l+1} - t^{l}}\mathbf{D}^l \mathbf{X}^l \overset{!}{=} \mathbf{F}^l(\mathbf{X}^l, \bm{u}^l, \bm t^l) \label{eq:lgr-colloc-const}
\end{align}
with the non-square differentiation matrix $\mathbf{D}^l\in \mathbb{R}^{N_\text{c} \times (N_\text{c}+1)}$ 
\begin{align*}
\mathbf{D}^l = \begin{pmatrix}\dot{{L}}_{1}(\tau_{1}) & \cdots & \dot{{L}}_{N_\text{c}}(\tau_{1}) & \dot{{L}}_{{N_\text{c}}+1}(\tau_{1})\\
\dot{{L}}_{1}(\tau_{2}) & \cdots & \dot{{L}}_{{N_\text{c}}}(\tau_{2}) & \dot{{L}}_{{N_\text{c}}+1}(\tau_{2})\\
& \vdots &\\
\dot{{L}}_{1}(\tau_{{N_\text{c}}}) & \cdots & \dot{{L}}_{{N_\text{c}}}(\tau_{{N_\text{c}}}) & \dot{{L}}_{{N_\text{c}}+1}(\tau_{{N_\text{c}}})
\end{pmatrix} 
\end{align*}
and the state and system-dynamics matrices
\begin{align}
  \begin{split}
  \mathbf{X}^l = \begin{pmatrix}
    (\bm x_1^l)^\text{T}\\
    \vdots\\
    (\bm x_{{N_\text{c}}}^l)^\text{T}\\[1em]
    (\bm x_{{N_\text{c}}+1}^l)^\text{T}
  \end{pmatrix}, \quad \mathbf{F}^l = \begin{pmatrix}
    \bm  f(\bm x_1^l,\bm u^l, t_1^l)^\text{T}\\
    \vdots\\
    \bm  f(\bm x_{N_\text{c}}^l,\bm u^l, t_{N_\text{c}}^l)^\text{T}\\
  \end{pmatrix}.\label{eq:lgrloc-X_F_matrices}
\end{split}\end{align}
The final node in each interval $\bm x^l_{N_\text{c}+1}$ is required to interpolate the state trajectory. No collocation constraint is imposed at this node, as reflected in $\mathbf{F}$. This property makes $\mathbf{D}$ non-square.

The state approximation order within each sub-interval is specified by $N_\text{c}$ and is assumed identical in all intervals. The number of intervals $N_\text{int}$ determines the number of control inputs, as shown in Fig.~\ref{fig:xuapproximation}. The parameter vector $\bm t$ defines the interval boundaries and hence the prediction horizon and mesh spacing.

\section{Automated Design and Evaluation Pipeline for Optimal Control Transcriptions (APOCT)}\label{sec:APOCT}

The goal of the APOCT is to optimize the transcription of a given MPC formulation according to custom design metrics, e.g., computation time or closed-loop performance. As input to the pipeline, we consider an MPC problem with a defined cost function~\eqref{eq:ocp_J}, system dynamics~\eqref{eq:ocp_x_dot}, and constraints~\eqref{eq:ocp_x_start}--\eqref{eq:ocp_x_terminal_region}. The formulation is transcribed into an NLP with the LGR-LOC method (Section~\ref{subsec:Sparsity-lgrloc}). We show how explicit and implicit formulations are derived from this formulation, leveraging the APOCT to evaluate both numerical approaches (Section~\ref{subsec:derivation_ms}). The LGR-LOC is configured by a set of parameters specifying its numerical order and mesh spacing. We employ Bayesian optimization (BO) to identify optimal parametrizations. This requires the definition of a sample space (Section~\ref{subsec:bo-sample-space}) and a custom multi-objective cost function (Section~\ref{subsec:bo-multi-obj}) that the BO seeks to minimize. As a result, the APOCT derives Pareto-optimal configurations for the MPC transcription that minimize the multi-objective metric. An overview of the APOCT is given in Fig.~\ref{fig:bo-pipeline}.

\subsection{Sparsity exploitation of LGR-LOC}\label{subsec:Sparsity-lgrloc}
We apply the proposed state and control input approximation of the LGR-LOC (Section~\ref{subsec:background}) to each interval $l\in \mathbb{I}_{[1,N_\text{int}]}$. The optimization variables are the states in each interval $\mathbf{X} = (\mathbf{X}^1,\ldots,\mathbf{X}^{N_\text{int}}, \bm x^{N_\text{int}+1})$ and the control inputs, one per interval, $\mathbf{U} = (\bm u^1,\ldots, \bm u^{N_\text{int}})$. The NLP is given as
\begin{subequations}\label{eq:lgrloc-nlp}
  \begin{align}
  \min_\mathbf{X,U}\quad& V_\text{f}(\bm x^{N_\text{int}+1}) + \sum\limits_{l=1}^{N_{\text{int}}}\frac{t^{l+1} - t^{l}}{2}\sum\limits_{i=1}^{{N_\text{c}}} w_i^l \mathcal{L}(\bm x_i^l, \bm u^l, t^l)\label{eq:lgr-loc-cost}   \\
  \text{s.t.} \quad& \frac{2}{t^{l+1} - t^{l}} \mathbf{D}^l \mathbf{X}^l = \mathbf{F}^l(\mathbf{X}^l, \bm{u}^l, \bm t^l), \,\, \forall l\in\mathbb{I}_{[1,N_\text{int}]}\label{eq:colloc-constraint}\\
  & \bm x_1^1 = \bm x_\text{start} \label{eq:lgrloc-initial}\\
  & \bm{x}^{l+1}_1 = \bm{x}^l_{N_\text{c}+1}, \quad \forall l\in\mathbb{I}_{[1,N_\text{int}]}\label{eq:lgr-loc-star-continuity}\\
  & \bm g(\mathbf X^1,\ldots, \mathbf X^{N_\text{int}}, \bm u^1, \ldots, \bm u^{N_\text{int}}) \leq 0\label{eq:general-constraints}\\
  & \bm x^{N_\text{int}+1}\in \mathcal{X}_\text{f}\label{eq:terminal-constraint}
  \end{align}
\end{subequations}
with the cost functional~\eqref{eq:lgr-loc-cost} approximated by Gauss-Radau quadrature with weights $w_i$~\cite{wang_symplectic_2021}, the dynamic constraints~\eqref{eq:colloc-constraint}, the initial condition~\eqref{eq:lgrloc-initial}, the continuity constraints~\eqref{eq:lgr-loc-star-continuity}, the path constraints~\eqref{eq:general-constraints}, and the terminal constraint~\eqref{eq:terminal-constraint} ensuring that the final state is within the terminal region $\mathcal{X}_\text{f}$.
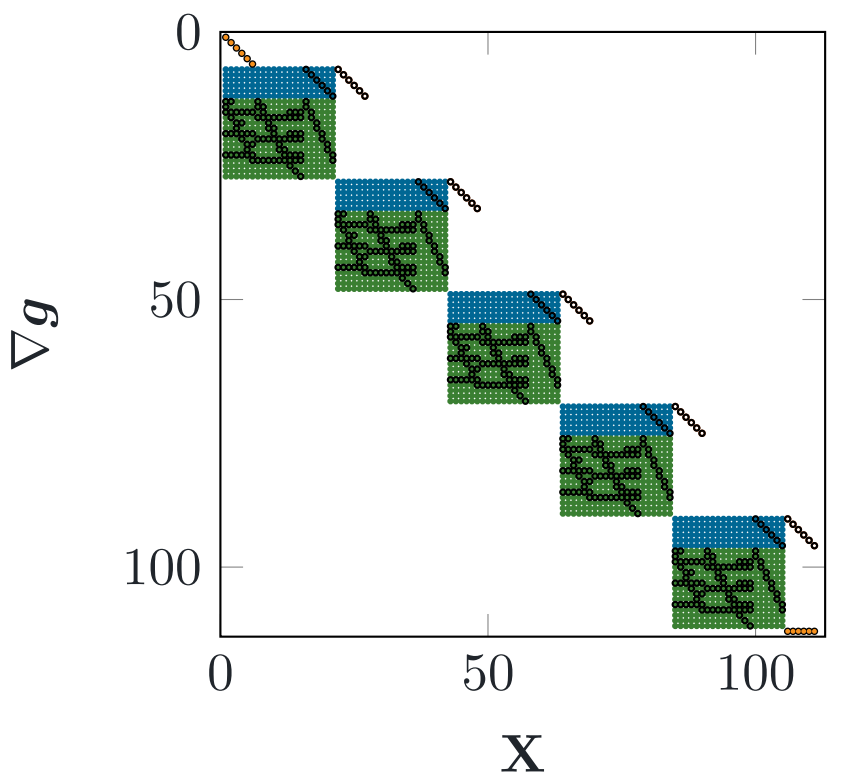
\begin{figure}[h]
  \centering
  \input{figures/lgrloc_sparsity/lgrloc_sparsity_ma.tex}
  \caption{Sparsity pattern for the LGR-LOC with $N_\text{int} = 5$ exploited by Fatrop~\cite{vanroye_fatrop_2023}. The constraints for collocating the dynamics~\eqref{eq:colloc-constraint} and path constraints~\eqref{eq:general-constraints} are shown in green, the gap-closing constraints~\eqref{eq:lgr-loc-star-continuity} in blue and the initial~\eqref{eq:lgrloc-initial} and terminal constraints~\eqref{eq:terminal-constraint} in orange.}\label{fig:sparsity-pattern}
\end{figure}

To ensure sparsity compatibility of the LGR-LOC with the nonlinear solver Fatrop~\cite{vanroye_fatrop_2023}, we introduce an additional state variable in each interval for the last interpolated state $\bm x_{N_\text{c}+1}^l$.
It represents the same state as the first node of the next interval $\bm x_1^{l+1}$. To ensure continuity, the gap-closing constraint~\eqref{eq:lgr-loc-star-continuity} is introduced. Figure~\ref{fig:sparsity-pattern} shows the sparsity pattern of the constraint Jacobian $\nabla g$ for an example with $N_\text{int} = 5$.


\subsection{Explicit and implicit formulations in the LGR-LOC}\label{subsec:derivation_ms}
In this section, we show that for a particular parameter choice the LGR-LOC~\eqref{eq:lgrloc-nlp} structurally includes a multiple-shooting formulation with a forward-Euler integrator. This property is then used in the APOCT pipeline.
\begin{proposition}\label{prop:lgrloc_parametrization}
	Consider the continuous-time OCP~\eqref{eq:continous-ocp}
	and its LGR-LOC discretization~\eqref{eq:lgrloc-nlp}, based on
	left-sided Legendre-Gauss-Radau collocation. The discretization
	is parameterized by the number of collocation points $N_\mathrm{c}^l$
	per interval, the number of intervals $N_\mathrm{int}$, and the
	interval boundaries $\bm t$.
	\par
	Then, for $N_\mathrm{c}^l = 1$ in every interval $l\in \mathbb{I}_{[1,N_{\text{int}}]}$, the collocation
	constraints are equivalent to the forward Euler integration
	equations. Consequently, after elimination of the interval endpoint variables, the resulting NLP is equivalent to a
	multiple-shooting formulation with a forward-Euler
	integrator.
	\par
	Moreover, for $N_\mathrm{c}^l > 1$ in every interval $l\in \mathbb{I}_{[1,N_{\text{int}}]}$, the LGR-LOC yields
	an implicit collocation discretization.
\end{proposition}

\begin{proof}
	Consider first $N_\mathrm{c}^l=1$ for all
	$l\in\mathbb{I}_{[1,N_\mathrm{int}]}$.
	The left-sided LGR scheme has the collocation point
	$\tau=-1$, yielding the differentiation matrix
	\begin{align}
	\mathbf D^l =
	\begin{pmatrix}
	-\frac{1}{2} & \frac{1}{2}
	\end{pmatrix}.
	\end{align}
	With the interval length $h^l=t^{l+1}-t^l$,
	the collocation constraints from
	Equation~\eqref{eq:lgr-colloc-const} become
	\begin{align}
	\frac{2}{h^l}
	\begin{pmatrix}
	-\frac{1}{2} & \frac{1}{2}
	\end{pmatrix}
	\begin{pmatrix}
	(\bm x_1^l)^\mathrm{T} \\
	(\bm x_2^l)^\mathrm{T}
	\end{pmatrix}
	=
	\bm f(\bm x_1^l,\bm u^l,t^l)^\mathrm{T}.
	\end{align}
	Rearranging yields
	\begin{align}
	\bm x_2^l
	=
	\bm x_1^l
	+
	h^l\bm f(\bm x_1^l,\bm u^l,t^l),
	\label{eq:app-hyper-MS}
	\end{align}
	which is exactly the forward Euler
	integration scheme.
	
	Combining Equation~\eqref{eq:app-hyper-MS} with the
	inter-interval continuity constraints~\eqref{eq:lgr-loc-star-continuity}
	$\bm x_1^{l+1}=\bm x_2^l$ yields
	\begin{align}
	\bm x_1^{l+1}
	=
	\bm x_1^l
	+
	h^l\bm f(\bm x_1^l,\bm u^l,t^l),
	\quad
	l\in\mathbb{I}_{[1,N_\mathrm{int}]}.
	\end{align}
	These are exactly the continuity constraints of a
	multiple-shooting formulation with a forward-Euler
	integrator. The final interval is propagated
	analogously to obtain the terminal state $\bm x^{N_\text{int}+1}$.
	
	Furthermore, the single LGR quadrature weight is
	$w_1=2$, such that the discretized stage cost reduces to
	\begin{align}
	\frac{h^l}{2}w_1
	\mathcal L(\bm x_1^l,\bm u^l,t^l)
	=
	h^l\mathcal L(\bm x_1^l,\bm u^l,t^l).
	\end{align}
	Thus, the objective also coincides with that of the
	corresponding multiple-shooting formulation employing
	left-rectangle quadrature. Consequently, the two
	NLP formulations are equivalent upon elimination of
	the interval endpoint variables.
	
	For $N_\mathrm{c}^l>1$, the additional LGR collocation
	points introduce intermediate state values
	into the collocation constraints. These values satisfy
	coupled implicit equations, yielding an implicit
	collocation formulation.
	This completes the proof.
\end{proof}

\begin{remark}
	As shown in Proposition~\ref{prop:lgrloc_parametrization},
	selecting $N_\mathrm{c}^l=1$ for all
	$l\in\mathbb{I}_{[1,N_\mathrm{int}]}$ yields a formulation
	equivalent, after elimination of the interval endpoint variables, to multiple shooting with a forward-Euler
	integrator. However, the LGR-LOC formulation retains the
	interval endpoint states $\bm{x}^l_{N_\text{c}+1}$
	as additional optimization variables
	in the NLP~\eqref{eq:lgrloc-nlp}, which can be eliminated
	in a native multiple-shooting implementation.
	Therefore, if the APOCT selects $N_\mathrm{c}=1$ for
	all intervals, a native multiple-shooting implementation
	may reduce computational overhead and improve efficiency.
\end{remark}

\subsection{Sample space}\label{subsec:bo-sample-space}
The APOCT requires a definition of a sample space specifying the set of all admissible LGR-LOC transcription configurations under investigation. We define the BO search space as
\begin{equation}\label{eq:bo-sample-space}
\mathcal{B}
:= \mathcal{C}_n \times \mathcal{I} \times \mathcal{T},
\end{equation}
where $\mathcal{C}_n$ is the set of admissible collocation approximation orders, $\mathcal{I}\subseteq \mathbb{I}_{>0}$ denotes the set of admissible number of intervals, and
$\mathcal{T} \subseteq \mathbb{R}_{>0}$ the admissible
values of the prediction horizon $T$.
Each sampled configuration~$\bm b = (N_\mathrm{c},N_{\mathrm{int}},T)$, with $\bm b \in \mathcal{B}$, defines an MPC transcription, with the mesh vector
$\bm{t}$ constructed from $(N_{\mathrm{int}},T)$
using the exponential coarsening strategy described below.

The admissible values of $N_{\mathrm{int}}$ depend on the control problem.
The parameter $N_c$ determines the approximation order and distinguishes between explicit and implicit transcriptions. To accommodate both, we define
\begin{equation}
\mathcal{C}_n := \{1,k_1,\ldots,k_n\},
\qquad
k_i \in \mathbb{I}, \quad k_i > 1,
\end{equation}
where $n \in \mathbb{I}$.

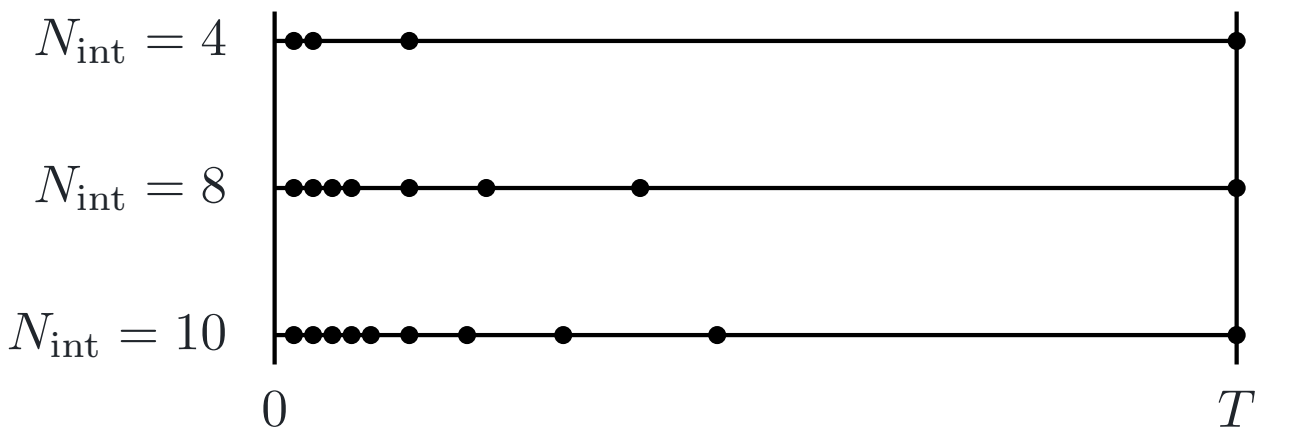
\begin{figure}
  \input{figures/t_sampling/t_sampling.tex}
  \caption{Interval spacing derived according to the exponential decay of sensitivity~\cite{shin_diffusinghorizon_2023}.\label{fig:spacing_t_bo}}
\end{figure}

The dimension of the mesh vector
$\bm{t}=(t^1,\ldots,t^{N_{\mathrm{int}}+1})$
depends on the sampled value of $N_{\mathrm{int}}$.
To obtain a fixed-dimensional parametrization of the prediction horizon we employ an exponential coarsening strategy inspired by the diffusing-horizon MPC approach of Shin and Zavala~\cite{shin_diffusinghorizon_2023} that defines the mapping\begin{equation}
\Phi_{\mathrm{exp}}:
(N_{\mathrm{int}},T)
\mapsto \bm{t}
\in \mathbb{R}^{N_{\mathrm{int}}+1},
\end{equation}
with
\begin{equation}
t^l = \max
\left\{
\begin{aligned}[c]
&(l-1) \cdot \Delta t_\text{MPC},\\
&\left\lfloor \left(\frac{T}{\Delta t_\text{MPC}}+1\right)^{\frac{l-2}{N_\text{int}}}\right\rfloor \Delta t_\text{MPC}
\end{aligned}
\right\}
\label{eq:coarsening-strategy}
\end{equation}
for $l=2,\ldots,N_{\mathrm{int}}$. The first interval starts at $t^1 = 0$ and the end time is $t^{N_\mathrm{int}+1} = T$. The resulting time grid contains a fixed number of points, with the spacing increasing toward the end of the horizon. Here, $\Delta t_\text{MPC}$ denotes the MPC sampling period. The grid construction is designed to respect the minimum temporal resolution and to retain a fine discretization near the beginning of the prediction horizon, where the system is most sensitive to future perturbations. For linear systems, the exponential coarsening strategy has been shown to preserve the exponential decay of sensitivity~\cite{shin_diffusinghorizon_2023}. Figure~\ref{fig:spacing_t_bo} illustrates the resulting
mesh spacing for different values of $N_{\mathrm{int}}$.

Recursive feasibility and stability of MPC have been established under suitable assumptions~\cite{rawlings_model_2017}. However, these proofs typically rely on shifting the previous solution, which is not generally feasible for nonuniform or adaptive discretizations, such as~\eqref{eq:coarsening-strategy}. Extending these guarantees to such transcriptions therefore requires further analysis.

\subsection{Multi-objective cost function}\label{subsec:bo-multi-obj}
A transcription structure for the LGR-LOC~\eqref{eq:lgrloc-nlp} is established with the properties described in the Proposition~\ref{prop:lgrloc_parametrization}. For the parametrization of the LGR-LOC, a sample space~\eqref{eq:bo-sample-space} is defined. Based on this, we deploy a BO with the multi-objective cost function
  \begin{align}
    \min_{\bm b \in \mathcal{B}}  \quad \bm \Pi = (\Pi_1(\bm b), \ldots, \Pi_K(\bm b)) \label{eq:bo-multiobjective}
  \end{align}
with $K\in \mathbb{I}_{>0}$ objectives. To minimize~\eqref{eq:bo-multiobjective}, the Tree-structured Parzen Estimator sampling method~\cite{ozaki_multiobjective_2020} in the Optuna framework~\cite{akiba_optuna_2019} is applied.

The objectives $\Pi_k$, $k=1,\ldots,K$, represent customized MPC performance metrics associated with the transcription, e.g., computation time and closed-loop cost. Minimization of the corresponding multi-objective cost function~\eqref{eq:bo-multiobjective} yields a set of Pareto optimal configurations~\cite{emmerich_tutorial_2018} for the MPC transcription.

The APOCT is terminated after a prespecified number of iterations. The user selects a transcription $\bm b \in \mathcal{B}$ according to individual performance preferences for the control problem under consideration. We propose a hierarchical selection procedure:  First, the numerical scheme (explicit or implicit and approximation order) $N_\text{c}$ is defined. Second, the number of intervals $N_\text{int}$ is evaluated, and as a last step, the prediction horizon $T$ and the interval spacing $\bm t$ according to~\eqref{eq:coarsening-strategy} is specified.

\section{Results for the APOCT in an On-Orbit Servicing scenario}\label{sec:results}
\begin{figure}[htb]
  \includegraphics[width=\linewidth]{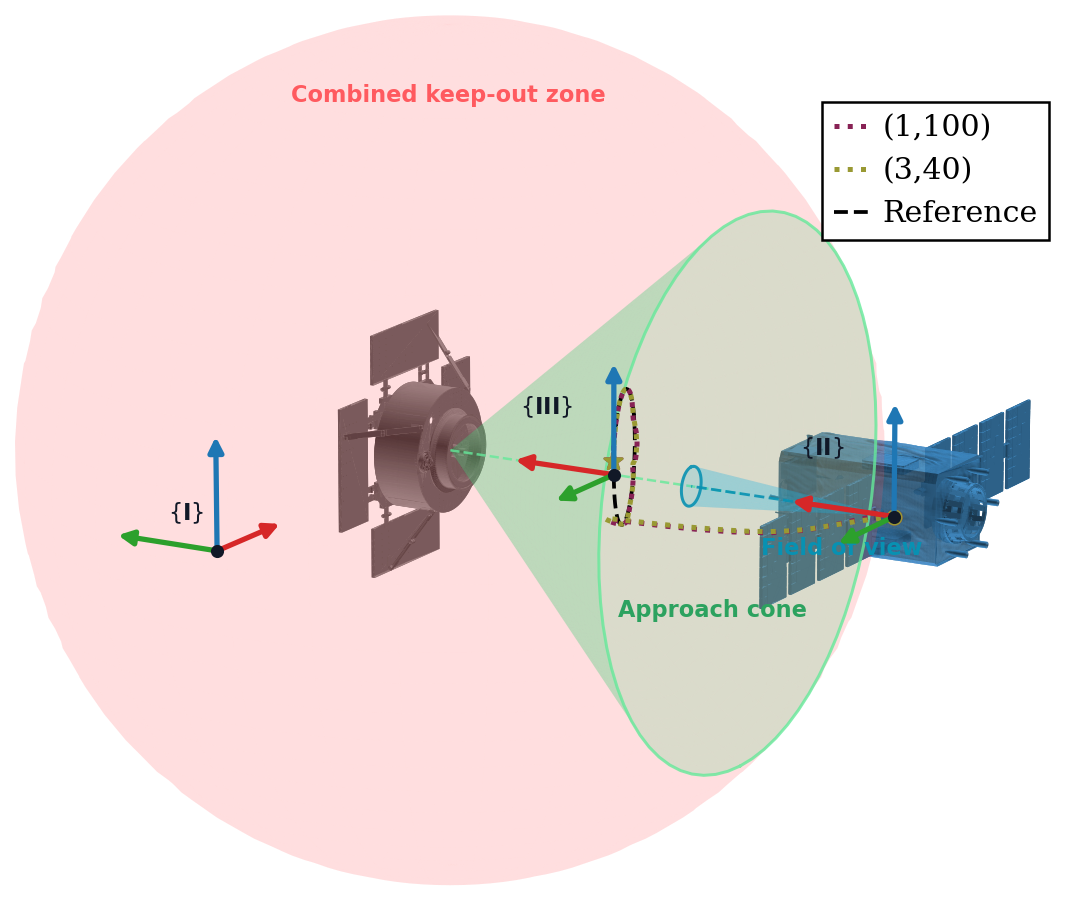}
  \caption{OOS scenario with the Servicer (right) and EuCROPIS Target (left). The motion planning trajectories for the transcription $\bm b_1 = (1,100)$ and $\bm b_2 = (3,40)$ in an Approach phase are shown.\label{fig:OOS-scenario}}
\end{figure}

The proposed APOCT is demonstrated on an OOS scenario with a Servicer and a Target satellite~\cite{kotting_esagnc_2026}, where we use EuCROPIS~\cite{hauslage_eu_2018} as the Target, shown in Fig.~\ref{fig:OOS-scenario}. 

The goal is to optimize the transcription of an existing six-degree-of-freedom (6-DoF) spacecraft MPC~\cite{kotting_esagnc_2026}. This MPC jointly controls the relative position and orientation of the Servicer while accounting for field-of-view, keep-out-zone, and approach-cone constraints, which results in a non-convex formulation. The MPC transcription is optimized in the context of open-loop motion planning and closed-loop MPC. The latter requires high-fidelity simulation for validation. This simulation is implemented in Modelica and exported as a Functional Mock-up Unit (FMU) in a Python environment~\cite{kotting_esagnc_2026}.

Figure~\ref{fig:OOS-scenario} gives an overview of the OOS scenario showing an exemplary predicted trajectory and maneuver constraints of the MPC. Furthermore, the different mission phases of an OOS maneuver~\cite{kotting_esagnc_2026} are represented. In this paper, the mission phases Rendezvous (frame \{\textbf{I}\}-\{\textbf{II}\}), Approach (frame \{\textbf{II}\}-\{\textbf{III}\}) and Tracking (frame \{\textbf{III}\}) are considered. 

\subsection{APOCT for motion planning application example}\label{sec:results-guidance}
The proposed APOCT is applied to a motion planning example. A single open-loop solution is computed without a warm-start solution for the approach phase (frame \{\textbf{II}\}-\{\textbf{III}\}). The prediction horizon for the motion planning scenario is constant and assumed to be $T=90$ s (full rotational period of the Target). Equidistant spacing is used. This reduces the parameter space to $\bm b = (N_\text{c}, N_\text{int}) \in \mathcal{B}_\text{OL}$ with
$\mathcal{B}_\text{OL} \mathrel{:=} \mathbb{I}_{[1,5]} \times \mathbb{I}_{[10,100]}$.

The multi-objective cost function~\eqref{eq:bo-multiobjective} with $K=2$ consists of $\mathbf{\Pi} = (\Pi_{t},\Pi_{\text{OL}})$, with $\Pi_{t}$ being the computation time of an open-loop iteration and $\Pi_{\text{OL}}$ the open-loop cost of the computed motion planning trajectory. This is evaluated by
\begin{align}
	\Pi_\text{OL}(\bm b) = \sum_{i} &(\bm x^*(t_i, \bm b)-\bm x_\text{ref}(t_i))^\mathsf{T} \mathbf{Q} \left(\bm x^*(t_i, \bm b)-\bm x_\text{ref}(t_i)\right)\notag\\ 
	&+ \bm u^*(t_i, \bm b)^\mathsf{T} \mathbf{R} \bm u^*(t_i, \bm b)\label{eq:bo-ol-cost}
\end{align}
with $\mathbf{Q,R}$ being the identical MPC weighting matrices from the Lagrangian $\mathcal{L}$ in~\eqref{eq:lgr-loc-cost}, $\bm x_\text{ref}$ being the propagated target pose evaluated at the time $t_i$, and $\bm x^*(t_i, \bm b), \bm u^*(t_i,\bm b)$ being the open-loop state and control-input trajectories generated with the transcription configuration $\bm b \in \mathcal{B}_\text{OL}$ evaluated at time $t_i$. The cost function~\eqref{eq:bo-ol-cost} is evaluated at 1000 uniformly spaced timestamps.

\begin{figure}[htb]
  \includegraphics[scale=1]{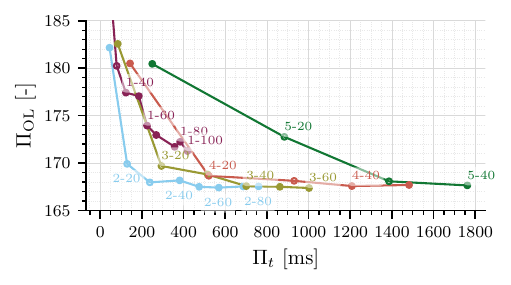}
  \caption{Pareto-optimal MPC transcriptions for Motion Planning derived by APOCT. The labels represent the tuples $\bm b = (N_\text{c},N_\text{int})$. All configurations $(1,\cdot)$ represent explicit formulations, while all labels $(\ge 2,\cdot)$ represent implicit formulations with increasing order. The second number is the number of subdomains.\label{fig:results-guidance}}
\end{figure}

The results are shown in Fig.~\ref{fig:results-guidance}. The explicit transcription method $N_\text{c} = 1$ achieves higher open-loop cost at comparable computation time than collocation transcriptions $N_\text{c} \geq 2$. All collocation transcriptions converge to a similar value of $\Pi_\text{OL}$ as the number of intervals increases.

In this exemplary motion planning case, collocation transcriptions with two or three collocation points represent a good trade-off between computation time and open-loop cost. Two trajectories $\bm b_1=(1,100)$ and $\bm b_2=(3,40)$ are represented in Fig.~\ref{fig:OOS-scenario}.

\subsection{APOCT for MPC application example}
The APOCT is applied to the closed-loop MPC example with variable prediction horizon. The mission phases for Rendezvous (RDV) (frame \{\textbf{I}\}-\{\textbf{II}\}), Approach (APP) (frame \{\textbf{II}\}-\{\textbf{III}\}) and Tracking (TRA) (frame \{\textbf{III}\}) are considered. The multi-objective cost function~\eqref{eq:bo-multiobjective} consists of $\mathbf{\Pi} = (\Pi_t, \Pi_\text{RDV/APP}, \Pi_\text{TRA})$. The computation time, measured as the 95th percentile over all MPC executions, is denoted by $\Pi_t$, the closed-loop performance for the sequence of a RDV and a APP phase as $\Pi_\text{RDV/APP}$, and the closed-loop performance for the TRA phase as $\Pi_\text{TRA}$. We distinguish between $\Pi_\text{RDV/APP}$ and $\Pi_\text{TRA}$ to account for the different structure of the control problem. In the case of RDV/APP, a step-response control problem is obtained, while TRA corresponds to a reference-tracking control problem. For the different use cases, separate simulations are executed in each BO iteration. The closed-loop sample space for the APOCT is $\bm b =(N_\text{c}, N_\text{int}, T) \in \mathcal{B}_\text{CL} := \mathbb{I}_{[1,3]} \times \mathbb{I}_{[4,10]} \times \mathbb{I}_{[2,10]}$.

The closed-loop performance is computed as
\begin{align}
	\Pi_m = \sum_{i} &(\bm x(t_i, \bm b)-\bm x_\text{ref}(t_i))^\mathsf{T} \mathbf{Q} \left(\bm x(t_i, \bm b)-\bm x_\text{ref}(t_i)\right)\notag\\ 
	&+ \bm u(t_i, \bm b)^\mathsf{T} \mathbf{R} \bm u(t_i, \bm b)\label{eq:bo-cl-cost}
\end{align}
with $m\in \{\text{RDV/APP, TRA}\}$. In contrast to motion planning, the state $\bm x(t_i,\bm b)$ represents the simulated closed-loop state of the spacecraft at time $t_i$ obtained by iteratively applying the first control input from each MPC solution $u(t_i,\bm b) = u_{t_i}^*(0, \bm b)$. Here, $u_{t_i}^*(t, \bm b)$ denotes the optimal control input solution trajectory from an MPC execution at time $t_i$. The MPC sampling period is $\Delta t_\text{MPC} = 0.1$ s.

To ease the selection of Pareto-optimal parametrizations $\bm b \in \mathcal{B}_\text{CL}$, the closed-loop performance for the different mission phases is combined into the cost
\begin{align*}
  J_\text{CL}(\bm b) =  \frac{1}{2}\left({\frac{\Pi_{\text{RDV/APP}}(\bm b)}{\max(\bm \Pi_{\text{RDV/APP}})} + \frac{\Pi_{\text{TRA}}(\bm b)}{\max(\bm \Pi_{\text{TRA}})}}\right).\label{eq:PiComb}
\end{align*}
 We denote this cost by $J_\text{CL}$ to distinguish it from an objective used in the BO~\eqref{eq:bo-multiobjective}.

\begin{figure}[htb]
  \includegraphics[scale=1]{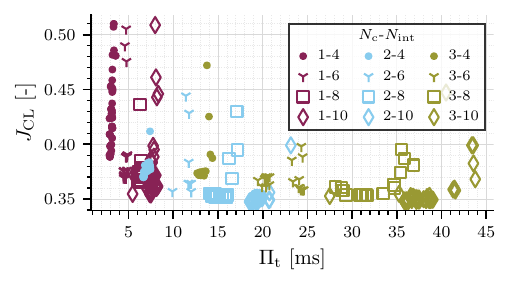}
  \caption{Pareto-optimal MPC transcriptions in the closed-loop scenario. All combinations $(1-\cdot)$ represent explicit MS transcriptions; all methods $(2-\cdot)$ and $(3-\cdot)$ represent implicit collocation transcriptions of order 2 and 3, respectively. The second number represents the number of intervals. Samples for the same method differ in the interval spacing $\bm t$.\label{fig:results-q95-PiComb}}
\end{figure}

The Pareto-optimal transcriptions are shown in Fig.~\ref{fig:results-q95-PiComb}. For the given OOS scenario, the explicit MS formulation with $N_\text{c} = 1$ is computationally the most efficient transcription. The closed-loop behavior degrades for a small number of intervals but improves as the number of intervals increases.

Implicit collocation methods with $N_\text{c} \geq 2$ are associated with higher computational costs than MS, while slight closed-loop performance improvements are achieved for an identical number of intervals. The selection of an optimal transcription parametrization $\bm b^* \in \mathcal{B}_\text{CL}$ from the Pareto front is characterized by a trade-off between computation time and closed-loop performance. Explicit multiple-shooting should be employed for minimal computation time, whereas collocation methods are preferred for optimal closed-loop performance. As shown in Fig.~\ref{fig:results-q95-PiComb}, a good trade-off between computation time and closed-loop performance can be achieved with the method $(1,6)$, i.e., by an explicit multiple-shooting formulation with 6 intervals.

\begin{figure}[htb]
  \centering
  \includegraphics[scale=1]{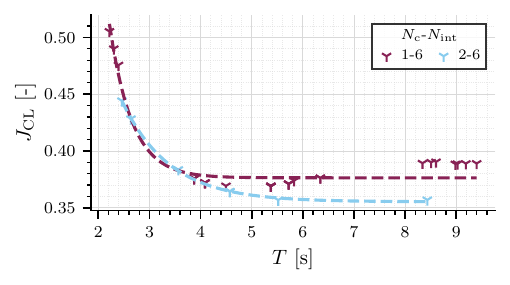}
  \caption{Influence of prediction horizon $T$ on the closed-loop cost.\label{fig:results-T-PiComb}}
\end{figure}

The influence of the prediction horizon $T$ for the selected transcription is shown in Fig.~\ref{fig:results-T-PiComb}. The exponential coarsening strategy~\eqref{eq:coarsening-strategy} is applied. For the preferred method $(1,6)$, the closed-loop cost decreases with increasing horizon length until it saturates. Exponential curves are fitted to the data samples. The collocation method $(2,6)$ with the identical number of intervals achieves improved $J_\text{CL}$. For the given OOS scenario, a suitable horizon length is $6$~s. This results in the optimized transcription configuration of $\bm{b}^* = (1, 6, 6)$.

\begin{figure}[htb]
  \includegraphics[scale=0.95]{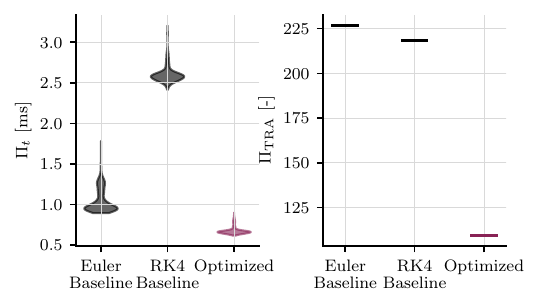}
  \caption{Comparison of the optimized MPC transcription with configuration $\bm b^* = (1,6,6)$ obtained via APOCT with the baseline formulation $\bm b_\text{baseline}=(1,10,1)$ using Euler and fourth-order Runge-Kutta integration. It shows the computation time $\Pi_t$ and the closed-loop cost $\Pi_\text{TRA}$ relative to the baseline MPC in a tracking scenario.\label{fig:baseline-lgrloc*}}
\end{figure}

Fig.~\ref{fig:baseline-lgrloc*} compares an optimized transcription derived via the APOCT configuration $\bm b^*$ with a baseline MPC using MS transcription for the Tracking phase with $N_\text{int} = 10$, $T=1$ s, and equidistant interval spacing of 0.1 s. The baseline results are shown for Euler forward integration and explicit fourth-order Runge-Kutta (RK4) integration. For Euler integration, the median computation time is reduced by more than 30~\% while the closed-loop cost is reduced by 50~\% in an OOS tracking mission phase. The combination of a longer prediction horizon and fewer optimization variables results in improved performance. The RK4 integration consumes more computation time, while the closed-loop performance is approximately 5\,\% better than with Euler integration.

\section{CONCLUSION}

We propose an automated design and evaluation pipeline for optimal control transcriptions. The pipeline derives Pareto-optimal MPC transcriptions that minimize a custom, multi-objective design metric, e.g., computation time and closed-loop performance. The APOCT transcribes the OCP into an NLP using a Legendre-Gauss-Radau method defined on multiple sub-intervals. We show how this method incorporates explicit and implicit numerical schemes, leveraging the pipeline to evaluate both numerical schemes. Furthermore, the LGR-LOC is parameterized by the order of approximation and by the local mesh structure, which accounts for the number of subdomains and the corresponding domain lengths.

The proposed APOCT is demonstrated on On-Orbit Servicing maneuvers for motion planning and closed-loop MPC. While for motion planning implicit collocation methods are found to offer the better trade-off, for closed-loop MPC an explicit multiple-shooting formulation is identified to be more efficient. Compared to a baseline formulation, the optimized transcription contains a reduced number of optimization variables, but an extended prediction horizon. This reduces computation time while improving closed-loop cost.

The strong interdependence between the optimal control transcription and the MPC problem is highlighted by the two examples. With the proposed pipeline, we derived a strategy that tailors MPC transcriptions to the specific control problem.

\addtolength{\textheight}{-9.3cm}   


\section*{ACKNOWLEDGMENT}
We thank the DLR Institute of Space Systems for providing us with a CAD model of the EuCROPIS satellite, shown in Fig.~\ref{fig:OOS-scenario}. Furthermore, we would like to thank Roberto Lampariello for helpful comments and annotations on the internal review version of this paper.



\bibliographystyle{IEEEtran}
\bibliography{references}   

\end{document}

%% file: packages.tex
\usepackage{amsmath}
\usepackage{amssymb}
\usepackage{bm}
\usepackage{graphicx}
\usepackage{tikz}
\usetikzlibrary{trees, positioning, shapes.geometric, fit}
\usepackage{pgfplots}
\usepackage[dvipsnames]{xcolor}
\usepackage{todonotes}
\usepackage{siunitx}
\usepackage{nicefrac}
\usepackage{cite}     
\makeatletter                 
\let\NAT@parse\undefined
\makeatother
\usepackage[hidelinks]{hyperref}       

\newtheorem{proposition}{Proposition}
\newtheorem{remark}{Remark}

\usetikzlibrary{arrows.meta,calc,positioning}

\definecolor{textcol}{HTML}{20252B}
\definecolor{muted}{HTML}{5C6975}
\definecolor{linecol}{HTML}{314252}
\definecolor{bluepanel}{HTML}{DCEBFA}
\definecolor{bluebox}{HTML}{F7FBFF}
\definecolor{bluedark}{HTML}{2F5F8F}
\definecolor{greenpanel}{HTML}{FAFFFA}
\definecolor{greenbox}{HTML}{DFF0E2}
\definecolor{greendark}{HTML}{2F7D4A}
\definecolor{orangebox}{HTML}{FCE7C9}
\definecolor{orangedark}{HTML}{B36A00}
\definecolor{purplebox}{HTML}{E8E0F4}
\definecolor{purpledark}{HTML}{6B4C9A}
\definecolor{redbox}{HTML}{F6D7D7}
\definecolor{reddark}{HTML}{9B3A3A}
\definecolor{greybox}{HTML}{F2F4F7}
\definecolor{greydark}{HTML}{53606D}

\tikzset{
  every node/.style={font=\sffamily, text=textcol},
  panel/.style args={#1/#2}{draw=#1, fill=#2, line width=0.85pt, rounded corners=5pt, inner sep=0pt},
  box/.style args={#1/#2}{draw=#1, fill=#2, line width=0.75pt, rounded corners=4pt, align=center, inner sep=2pt},
  arrow/.style={-{Latex[length=2.3mm,width=1.7mm]}, line width=0.85pt, draw=#1},
  arrow/.default=linecol
}

%% file: figures/schematic_methodology/schematic_methodology.tex


\begin{tikzpicture}

\node[panel=bluedark/bluepanel, minimum width=4.45cm, minimum height=1.65cm]
  (universalpanel) at (6.05, 6.55) {};

\node[box=bluedark/bluebox, minimum width=1.2cm, minimum height=0.50cm,
      font=\scriptsize, text width=0.6cm,
      at=(universalpanel.center), xshift=-1.5cm, yshift=-0.4cm] (nodeNl)
  {$N_\text{c}$};

\node[box=bluedark/bluebox, minimum width=1.2cm, minimum height=0.50cm,
      font=\scriptsize, text width=0.6cm,
      at=(universalpanel.center), xshift=-0.0cm, yshift=-0.4cm] (nodeNint)
  {$N_\text{int}$};

  \node[box=bluedark/bluebox, minimum width=1.2cm, minimum height=0.50cm,
      font=\scriptsize, text width=0.6cm,
      at=(universalpanel.center), xshift=1.5cm, yshift=-0.4cm] (nodeT)
  {$\bm t$};





\node[font=\footnotesize\bfseries, text=bluedark,
text width=4cm, below=0.18cm of universalpanel.north, align=center] (unilabel)
{Generalized MPC transcription (LGR-LOC) \tiny (Section \ref{subsec:Sparsity-lgrloc}) \& \ref{subsec:derivation_ms}};

\node[box=black/white, minimum width=2.80cm, minimum height=0.8cm,
      font=\footnotesize\bfseries, text width=2.55cm, right=1cm of universalpanel.east]
  (problem)
  {MPC problem};
\draw[arrow=greydark] (problem.west) -- (universalpanel.east);

\node[panel=greendark/greenpanel, minimum width=4.35cm, minimum height=3.90cm, below=0.5cm of universalpanel.south]
  (bopanel) {};

\node[font=\footnotesize\bfseries, text=greendark,
      below=0.18cm of bopanel.north, align=center] (bolabel)
  {Bayesian Optimization (BO)};

\node[box=greendark/greenbox, minimum width=4cm, minimum height=0.68cm,
      font=\scriptsize\bfseries, text width=3.5cm,
      below=0.1cm of bolabel] (sample)
  {1\quad LGR-LOC parametrization (\tiny Section~\ref{subsec:bo-sample-space})};


\node[box=greendark/greenbox, minimum width=4cm, minimum height=0.68cm,
      font=\scriptsize\bfseries, text width=3.5cm,
      below=0.43cm of sample] (evaluate)
  {2\quad Multi-objective optimization (\tiny Section~\ref{subsec:bo-multi-obj})};

\node[box=greendark/greenbox, minimum width=4cm, minimum height=0.68cm,
      font=\scriptsize\bfseries, text width=3.5cm,
      below=0.43cm of evaluate] (analyse)
  {3\quad Pareto optimal MPC transcription (\tiny Section~\ref{subsec:bo-multi-obj})};

\draw[arrow=greendark]  (sample)      -- (evaluate);
\node[font=\scriptsize, align=left, text=greendark, anchor=west, 
      below=0.06cm of sample.south, xshift=0.6cm, inner sep=2pt] (simulatedescription)
  {simulate};
\draw[arrow=greendark] (evaluate)    -- (analyse);

\draw[arrow=greendark]
  (evaluate.east) .. controls
    ([xshift=0.75cm]evaluate.east) and ([xshift=0.75cm]sample.east) ..
  (sample.east);

\node[font=\scriptsize, align=left, text=greendark, anchor=west, 
      right=1.7cm of simulatedescription.east, fill=white, inner sep=2pt]
  {iterate};

\draw[arrow=bluedark] (universalpanel.south) -- (bopanel.north);

\node[
  box=black/white,
  minimum width=2.80cm,
  minimum height=0.8cm,
  font=\footnotesize\bfseries,
  text width=2.55cm,
  anchor=west,
  at={([xshift=1cm]bopanel.east |- analyse.east)}
] (tailored)
{Optimized MPC transcription};

\draw[arrow=greendark] (analyse.east) -- (tailored.west);
\node[font=\scriptsize, align=left, text=greendark, anchor=west, 
      right=0.05cm of analyse.east, yshift=0.2cm, fill=white, inner sep=2pt]
  {select};
\end{tikzpicture}

%% file: figures/lgrloc_nodes_schematic/ps_lgl_lgr_plot_loc.tex


\begin{tikzpicture}[scale=3, xscale=0.5]

\draw[thick,->] (-1, -0.5) -- (3.3, -0.5); 
\draw[thick, ->] (-1, -0.5) -- (-1, -0.1); 

\draw[thick,  black, dashed] (-1, -0.7) -- (-1, 0); 
\draw[thick,  black, dashed] (1, -0.7) -- (1, 0); 
\draw[thick,  black, dashed] (3, -0.7) -- (3, 0); 

\node[font=\small] at (3.3, -0.4) {$t$}; 
\node[font=\small] at (-1.2, 0.05) {$w$}; 
\node[font=\small] at (-1.75, -0.5) {LGR-LOC}; 

\node[fill=white,font=\small] at (-1, -0.8) {$t^l$};
\node[fill=white,font=\small] at (1, -0.8) {$t^{l+1}$};
\node[fill=white,font=\small] at (3, -0.8) {$t^{l+2}$};



\draw[thick, ycomb, yshift=-0.5cm, yscale=0.5] plot coordinates {
(-1.000000, 0.08)
(-0.720480, 0.446208)
(-0.167181, 0.623653)
(0.446314, 0.562712)
(0.885792, 0.287427)
};

\draw[thick, ycomb, yshift=-0.5cm, shift={(2,0)}, yscale=0.5] plot coordinates {
(-1.000000, 0.08)
(-0.720480, 0.446208)
(-0.167181, 0.623653)
(0.446314, 0.562712)
(0.885792, 0.287427)
};



\foreach \x/\y in {-1.0000/0.0800, -0.7205/0.4462, -0.1672/0.6237, 0.4463/0.5627, 0.8858/0.2874} {
    \node[fill=black, rectangle, inner sep=2pt] at (\x, -0.5) {};  
}

\foreach \x/\y in {-1.0000/0.0800, -0.7205/0.4462, -0.1672/0.6237, 0.4463/0.5627, 0.8858/0.2874} {
    \node[fill=black, rectangle, inner sep=2pt,shift={(3,0)}] at (\x, -0.5) {};  
}

\end{tikzpicture}


%% file: figures/lgrloc_sparsity/lgrloc_sparsity_ma.tex
\definecolor{mycolor1}{rgb}{0.00000,1.00000,1.00000}%
%

\begin{tikzpicture}[baseline=(current bounding box.center)]

\begin{axis}[%
width=4cm,
height=4cm,
at={(0.0,0)},
scale only axis,
xmin=0,
xmax=113,
xlabel style={font=\color{white!15!black}},
y dir=reverse,
ymin=0,
ymax=113,
xlabel=$\mathbf{X}$,
ylabel=$\nabla \bm g$,
axis background/.style={fill=white},
legend style={legend cell align=left, align=left, draw=white!15!black}
]
\addplot [color=MidnightBlue  , draw=none, mark size=0.4pt, mark=*, mark options={solid, fill=white, MidnightBlue  }]
  table[row sep=crcr]{%
1	7\\
1	8\\
1	9\\
1	10\\
1	11\\
1	12\\
2	7\\
2	8\\
2	9\\
2	10\\
2	11\\
2	12\\
3	7\\
3	8\\
3	9\\
3	10\\
3	11\\
3	12\\
4	7\\
4	8\\
4	9\\
4	10\\
4	11\\
4	12\\
5	7\\
5	8\\
5	9\\
5	10\\
5	11\\
5	12\\
6	7\\
6	8\\
6	9\\
6	10\\
6	11\\
6	12\\
22	28\\
22	29\\
22	30\\
22	31\\
22	32\\
22	33\\
23	28\\
23	29\\
23	30\\
23	31\\
23	32\\
23	33\\
24	28\\
24	29\\
24	30\\
24	31\\
24	32\\
24	33\\
25	28\\
25	29\\
25	30\\
25	31\\
25	32\\
25	33\\
26	28\\
26	29\\
26	30\\
26	31\\
26	32\\
26	33\\
27	28\\
27	29\\
27	30\\
27	31\\
27	32\\
27	33\\
43	49\\
43	50\\
43	51\\
43	52\\
43	53\\
43	54\\
44	49\\
44	50\\
44	51\\
44	52\\
44	53\\
44	54\\
45	49\\
45	50\\
45	51\\
45	52\\
45	53\\
45	54\\
46	49\\
46	50\\
46	51\\
46	52\\
46	53\\
46	54\\
47	49\\
47	50\\
47	51\\
47	52\\
47	53\\
47	54\\
48	49\\
48	50\\
48	51\\
48	52\\
48	53\\
48	54\\
64	70\\
64	71\\
64	72\\
64	73\\
64	74\\
64	75\\
65	70\\
65	71\\
65	72\\
65	73\\
65	74\\
65	75\\
66	70\\
66	71\\
66	72\\
66	73\\
66	74\\
66	75\\
67	70\\
67	71\\
67	72\\
67	73\\
67	74\\
67	75\\
68	70\\
68	71\\
68	72\\
68	73\\
68	74\\
68	75\\
69	70\\
69	71\\
69	72\\
69	73\\
69	74\\
69	75\\
85	91\\
85	92\\
85	93\\
85	94\\
85	95\\
85	96\\
86	91\\
86	92\\
86	93\\
86	94\\
86	95\\
86	96\\
87	91\\
87	92\\
87	93\\
87	94\\
87	95\\
87	96\\
88	91\\
88	92\\
88	93\\
88	94\\
88	95\\
88	96\\
89	91\\
89	92\\
89	93\\
89	94\\
89	95\\
89	96\\
90	91\\
90	92\\
90	93\\
90	94\\
90	95\\
90	96\\
7	7\\
7	8\\
7	9\\
7	10\\
7	11\\
7	12\\
8	7\\
8	8\\
8	9\\
8	10\\
8	11\\
8	12\\
9	7\\
9	8\\
9	9\\
9	10\\
9	11\\
9	12\\
10	7\\
10	8\\
10	9\\
10	10\\
10	11\\
10	12\\
11	7\\
11	8\\
11	9\\
11	10\\
11	11\\
11	12\\
12	7\\
12	8\\
12	9\\
12	10\\
12	11\\
12	12\\
13	7\\
13	8\\
13	9\\
13	10\\
13	11\\
13	12\\
14	7\\
14	8\\
14	9\\
14	10\\
14	11\\
14	12\\
15	7\\
15	8\\
15	9\\
15	10\\
15	11\\
15	12\\
16	7\\
16	8\\
16	9\\
16	10\\
16	11\\
16	12\\
17	7\\
17	8\\
17	9\\
17	10\\
17	11\\
17	12\\
18	7\\
18	8\\
18	9\\
18	10\\
18	11\\
18	12\\
19	7\\
19	8\\
19	9\\
19	10\\
19	11\\
19	12\\
20	7\\
20	8\\
20	9\\
20	10\\
20	11\\
20	12\\
21	7\\
21	8\\
21	9\\
21	10\\
21	11\\
21	12\\
28	28\\
28	29\\
28	30\\
28	31\\
28	32\\
28	33\\
29	28\\
29	29\\
29	30\\
29	31\\
29	32\\
29	33\\
30	28\\
30	29\\
30	30\\
30	31\\
30	32\\
30	33\\
31	28\\
31	29\\
31	30\\
31	31\\
31	32\\
31	33\\
32	28\\
32	29\\
32	30\\
32	31\\
32	32\\
32	33\\
33	28\\
33	29\\
33	30\\
33	31\\
33	32\\
33	33\\
34	28\\
34	29\\
34	30\\
34	31\\
34	32\\
34	33\\
35	28\\
35	29\\
35	30\\
35	31\\
35	32\\
35	33\\
36	28\\
36	29\\
36	30\\
36	31\\
36	32\\
36	33\\
37	28\\
37	29\\
37	30\\
37	31\\
37	32\\
37	33\\
38	28\\
38	29\\
38	30\\
38	31\\
38	32\\
38	33\\
39	28\\
39	29\\
39	30\\
39	31\\
39	32\\
39	33\\
40	28\\
40	29\\
40	30\\
40	31\\
40	32\\
40	33\\
41	28\\
41	29\\
41	30\\
41	31\\
41	32\\
41	33\\
42	28\\
42	29\\
42	30\\
42	31\\
42	32\\
42	33\\
49	49\\
49	50\\
49	51\\
49	52\\
49	53\\
49	54\\
50	49\\
50	50\\
50	51\\
50	52\\
50	53\\
50	54\\
51	49\\
51	50\\
51	51\\
51	52\\
51	53\\
51	54\\
52	49\\
52	50\\
52	51\\
52	52\\
52	53\\
52	54\\
53	49\\
53	50\\
53	51\\
53	52\\
53	53\\
53	54\\
54	49\\
54	50\\
54	51\\
54	52\\
54	53\\
54	54\\
55	49\\
55	50\\
55	51\\
55	52\\
55	53\\
55	54\\
56	49\\
56	50\\
56	51\\
56	52\\
56	53\\
56	54\\
57	49\\
57	50\\
57	51\\
57	52\\
57	53\\
57	54\\
58	49\\
58	50\\
58	51\\
58	52\\
58	53\\
58	54\\
59	49\\
59	50\\
59	51\\
59	52\\
59	53\\
59	54\\
60	49\\
60	50\\
60	51\\
60	52\\
60	53\\
60	54\\
61	49\\
61	50\\
61	51\\
61	52\\
61	53\\
61	54\\
62	49\\
62	50\\
62	51\\
62	52\\
62	53\\
62	54\\
63	49\\
63	50\\
63	51\\
63	52\\
63	53\\
63	54\\
70	70\\
70	71\\
70	72\\
70	73\\
70	74\\
70	75\\
71	70\\
71	71\\
71	72\\
71	73\\
71	74\\
71	75\\
72	70\\
72	71\\
72	72\\
72	73\\
72	74\\
72	75\\
73	70\\
73	71\\
73	72\\
73	73\\
73	74\\
73	75\\
74	70\\
74	71\\
74	72\\
74	73\\
74	74\\
74	75\\
75	70\\
75	71\\
75	72\\
75	73\\
75	74\\
75	75\\
76	70\\
76	71\\
76	72\\
76	73\\
76	74\\
76	75\\
77	70\\
77	71\\
77	72\\
77	73\\
77	74\\
77	75\\
78	70\\
78	71\\
78	72\\
78	73\\
78	74\\
78	75\\
79	70\\
79	71\\
79	72\\
79	73\\
79	74\\
79	75\\
80	70\\
80	71\\
80	72\\
80	73\\
80	74\\
80	75\\
81	70\\
81	71\\
81	72\\
81	73\\
81	74\\
81	75\\
82	70\\
82	71\\
82	72\\
82	73\\
82	74\\
82	75\\
83	70\\
83	71\\
83	72\\
83	73\\
83	74\\
83	75\\
84	70\\
84	71\\
84	72\\
84	73\\
84	74\\
84	75\\
91	91\\
91	92\\
91	93\\
91	94\\
91	95\\
91	96\\
92	91\\
92	92\\
92	93\\
92	94\\
92	95\\
92	96\\
93	91\\
93	92\\
93	93\\
93	94\\
93	95\\
93	96\\
94	91\\
94	92\\
94	93\\
94	94\\
94	95\\
94	96\\
95	91\\
95	92\\
95	93\\
95	94\\
95	95\\
95	96\\
96	91\\
96	92\\
96	93\\
96	94\\
96	95\\
96	96\\
97	91\\
97	92\\
97	93\\
97	94\\
97	95\\
97	96\\
98	91\\
98	92\\
98	93\\
98	94\\
98	95\\
98	96\\
99	91\\
99	92\\
99	93\\
99	94\\
99	95\\
99	96\\
100	91\\
100	92\\
100	93\\
100	94\\
100	95\\
100	96\\
101	91\\
101	92\\
101	93\\
101	94\\
101	95\\
101	96\\
102	91\\
102	92\\
102	93\\
102	94\\
102	95\\
102	96\\
103	91\\
103	92\\
103	93\\
103	94\\
103	95\\
103	96\\
104	91\\
104	92\\
104	93\\
104	94\\
104	95\\
104	96\\
105	91\\
105	92\\
105	93\\
105	94\\
105	95\\
105	96\\
};

\addplot [color=OliveGreen , draw=none, mark size=0.4pt, mark=*, mark options={solid, fill=white, OliveGreen }]
  table[row sep=crcr]{%
1	13\\
1	14\\
1	15\\
1	16\\
1	17\\
1	18\\
1	19\\
1	20\\
1	21\\
1	22\\
1	23\\
1	24\\
1	25\\
1	26\\
1	27\\
2	13\\
2	14\\
2	15\\
2	16\\
2	17\\
2	18\\
2	19\\
2	20\\
2	21\\
2	22\\
2	23\\
2	24\\
2	25\\
2	26\\
2	27\\
3	13\\
3	14\\
3	15\\
3	16\\
3	17\\
3	18\\
3	19\\
3	20\\
3	21\\
3	22\\
3	23\\
3	24\\
3	25\\
3	26\\
3	27\\
4	13\\
4	14\\
4	15\\
4	16\\
4	17\\
4	18\\
4	19\\
4	20\\
4	21\\
4	22\\
4	23\\
4	24\\
4	25\\
4	26\\
4	27\\
5	13\\
5	14\\
5	15\\
5	16\\
5	17\\
5	18\\
5	19\\
5	20\\
5	21\\
5	22\\
5	23\\
5	24\\
5	25\\
5	26\\
5	27\\
6	13\\
6	14\\
6	15\\
6	16\\
6	17\\
6	18\\
6	19\\
6	20\\
6	21\\
6	22\\
6	23\\
6	24\\
6	25\\
6	26\\
6	27\\
22	34\\
22	35\\
22	36\\
22	37\\
22	38\\
22	39\\
22	40\\
22	41\\
22	42\\
22	43\\
22	44\\
22	45\\
22	46\\
22	47\\
22	48\\
23	34\\
23	35\\
23	36\\
23	37\\
23	38\\
23	39\\
23	40\\
23	41\\
23	42\\
23	43\\
23	44\\
23	45\\
23	46\\
23	47\\
23	48\\
24	34\\
24	35\\
24	36\\
24	37\\
24	38\\
24	39\\
24	40\\
24	41\\
24	42\\
24	43\\
24	44\\
24	45\\
24	46\\
24	47\\
24	48\\
25	34\\
25	35\\
25	36\\
25	37\\
25	38\\
25	39\\
25	40\\
25	41\\
25	42\\
25	43\\
25	44\\
25	45\\
25	46\\
25	47\\
25	48\\
26	34\\
26	35\\
26	36\\
26	37\\
26	38\\
26	39\\
26	40\\
26	41\\
26	42\\
26	43\\
26	44\\
26	45\\
26	46\\
26	47\\
26	48\\
27	34\\
27	35\\
27	36\\
27	37\\
27	38\\
27	39\\
27	40\\
27	41\\
27	42\\
27	43\\
27	44\\
27	45\\
27	46\\
27	47\\
27	48\\
43	55\\
43	56\\
43	57\\
43	58\\
43	59\\
43	60\\
43	61\\
43	62\\
43	63\\
43	64\\
43	65\\
43	66\\
43	67\\
43	68\\
43	69\\
44	55\\
44	56\\
44	57\\
44	58\\
44	59\\
44	60\\
44	61\\
44	62\\
44	63\\
44	64\\
44	65\\
44	66\\
44	67\\
44	68\\
44	69\\
45	55\\
45	56\\
45	57\\
45	58\\
45	59\\
45	60\\
45	61\\
45	62\\
45	63\\
45	64\\
45	65\\
45	66\\
45	67\\
45	68\\
45	69\\
46	55\\
46	56\\
46	57\\
46	58\\
46	59\\
46	60\\
46	61\\
46	62\\
46	63\\
46	64\\
46	65\\
46	66\\
46	67\\
46	68\\
46	69\\
47	55\\
47	56\\
47	57\\
47	58\\
47	59\\
47	60\\
47	61\\
47	62\\
47	63\\
47	64\\
47	65\\
47	66\\
47	67\\
47	68\\
47	69\\
48	55\\
48	56\\
48	57\\
48	58\\
48	59\\
48	60\\
48	61\\
48	62\\
48	63\\
48	64\\
48	65\\
48	66\\
48	67\\
48	68\\
48	69\\
64	76\\
64	77\\
64	78\\
64	79\\
64	80\\
64	81\\
64	82\\
64	83\\
64	84\\
64	85\\
64	86\\
64	87\\
64	88\\
64	89\\
64	90\\
65	76\\
65	77\\
65	78\\
65	79\\
65	80\\
65	81\\
65	82\\
65	83\\
65	84\\
65	85\\
65	86\\
65	87\\
65	88\\
65	89\\
65	90\\
66	76\\
66	77\\
66	78\\
66	79\\
66	80\\
66	81\\
66	82\\
66	83\\
66	84\\
66	85\\
66	86\\
66	87\\
66	88\\
66	89\\
66	90\\
67	76\\
67	77\\
67	78\\
67	79\\
67	80\\
67	81\\
67	82\\
67	83\\
67	84\\
67	85\\
67	86\\
67	87\\
67	88\\
67	89\\
67	90\\
68	76\\
68	77\\
68	78\\
68	79\\
68	80\\
68	81\\
68	82\\
68	83\\
68	84\\
68	85\\
68	86\\
68	87\\
68	88\\
68	89\\
68	90\\
69	76\\
69	77\\
69	78\\
69	79\\
69	80\\
69	81\\
69	82\\
69	83\\
69	84\\
69	85\\
69	86\\
69	87\\
69	88\\
69	89\\
69	90\\
85	97\\
85	98\\
85	99\\
85	100\\
85	101\\
85	102\\
85	103\\
85	104\\
85	105\\
85	106\\
85	107\\
85	108\\
85	109\\
85	110\\
85	111\\
86	97\\
86	98\\
86	99\\
86	100\\
86	101\\
86	102\\
86	103\\
86	104\\
86	105\\
86	106\\
86	107\\
86	108\\
86	109\\
86	110\\
86	111\\
87	97\\
87	98\\
87	99\\
87	100\\
87	101\\
87	102\\
87	103\\
87	104\\
87	105\\
87	106\\
87	107\\
87	108\\
87	109\\
87	110\\
87	111\\
88	97\\
88	98\\
88	99\\
88	100\\
88	101\\
88	102\\
88	103\\
88	104\\
88	105\\
88	106\\
88	107\\
88	108\\
88	109\\
88	110\\
88	111\\
89	97\\
89	98\\
89	99\\
89	100\\
89	101\\
89	102\\
89	103\\
89	104\\
89	105\\
89	106\\
89	107\\
89	108\\
89	109\\
89	110\\
89	111\\
90	97\\
90	98\\
90	99\\
90	100\\
90	101\\
90	102\\
90	103\\
90	104\\
90	105\\
90	106\\
90	107\\
90	108\\
90	109\\
90	110\\
90	111\\
7	13\\
7	14\\
7	15\\
7	16\\
7	17\\
7	18\\
7	19\\
7	20\\
7	21\\
7	22\\
7	23\\
7	24\\
7	25\\
7	26\\
7	27\\
8	13\\
8	14\\
8	15\\
8	16\\
8	17\\
8	18\\
8	19\\
8	20\\
8	21\\
8	22\\
8	23\\
8	24\\
8	25\\
8	26\\
8	27\\
9	13\\
9	14\\
9	15\\
9	16\\
9	17\\
9	18\\
9	19\\
9	20\\
9	21\\
9	22\\
9	23\\
9	24\\
9	25\\
9	26\\
9	27\\
10	13\\
10	14\\
10	15\\
10	16\\
10	17\\
10	18\\
10	19\\
10	20\\
10	21\\
10	22\\
10	23\\
10	24\\
10	25\\
10	26\\
10	27\\
11	13\\
11	14\\
11	15\\
11	16\\
11	17\\
11	18\\
11	19\\
11	20\\
11	21\\
11	22\\
11	23\\
11	24\\
11	25\\
11	26\\
11	27\\
12	13\\
12	14\\
12	15\\
12	16\\
12	17\\
12	18\\
12	19\\
12	20\\
12	21\\
12	22\\
12	23\\
12	24\\
12	25\\
12	26\\
12	27\\
13	13\\
13	14\\
13	15\\
13	16\\
13	17\\
13	18\\
13	19\\
13	20\\
13	21\\
13	22\\
13	23\\
13	24\\
13	25\\
13	26\\
13	27\\
14	13\\
14	14\\
14	15\\
14	16\\
14	17\\
14	18\\
14	19\\
14	20\\
14	21\\
14	22\\
14	23\\
14	24\\
14	25\\
14	26\\
14	27\\
15	13\\
15	14\\
15	15\\
15	16\\
15	17\\
15	18\\
15	19\\
15	20\\
15	21\\
15	22\\
15	23\\
15	24\\
15	25\\
15	26\\
15	27\\
16	13\\
16	14\\
16	15\\
16	16\\
16	17\\
16	18\\
16	19\\
16	20\\
16	21\\
16	22\\
16	23\\
16	24\\
16	25\\
16	26\\
16	27\\
17	13\\
17	14\\
17	15\\
17	16\\
17	17\\
17	18\\
17	19\\
17	20\\
17	21\\
17	22\\
17	23\\
17	24\\
17	25\\
17	26\\
17	27\\
18	13\\
18	14\\
18	15\\
18	16\\
18	17\\
18	18\\
18	19\\
18	20\\
18	21\\
18	22\\
18	23\\
18	24\\
18	25\\
18	26\\
18	27\\
19	13\\
19	14\\
19	15\\
19	16\\
19	17\\
19	18\\
19	19\\
19	20\\
19	21\\
19	22\\
19	23\\
19	24\\
19	25\\
19	26\\
19	27\\
20	13\\
20	14\\
20	15\\
20	16\\
20	17\\
20	18\\
20	19\\
20	20\\
20	21\\
20	22\\
20	23\\
20	24\\
20	25\\
20	26\\
20	27\\
21	13\\
21	14\\
21	15\\
21	16\\
21	17\\
21	18\\
21	19\\
21	20\\
21	21\\
21	22\\
21	23\\
21	24\\
21	25\\
21	26\\
21	27\\
28	34\\
28	35\\
28	36\\
28	37\\
28	38\\
28	39\\
28	40\\
28	41\\
28	42\\
28	43\\
28	44\\
28	45\\
28	46\\
28	47\\
28	48\\
29	34\\
29	35\\
29	36\\
29	37\\
29	38\\
29	39\\
29	40\\
29	41\\
29	42\\
29	43\\
29	44\\
29	45\\
29	46\\
29	47\\
29	48\\
30	34\\
30	35\\
30	36\\
30	37\\
30	38\\
30	39\\
30	40\\
30	41\\
30	42\\
30	43\\
30	44\\
30	45\\
30	46\\
30	47\\
30	48\\
31	34\\
31	35\\
31	36\\
31	37\\
31	38\\
31	39\\
31	40\\
31	41\\
31	42\\
31	43\\
31	44\\
31	45\\
31	46\\
31	47\\
31	48\\
32	34\\
32	35\\
32	36\\
32	37\\
32	38\\
32	39\\
32	40\\
32	41\\
32	42\\
32	43\\
32	44\\
32	45\\
32	46\\
32	47\\
32	48\\
33	34\\
33	35\\
33	36\\
33	37\\
33	38\\
33	39\\
33	40\\
33	41\\
33	42\\
33	43\\
33	44\\
33	45\\
33	46\\
33	47\\
33	48\\
34	34\\
34	35\\
34	36\\
34	37\\
34	38\\
34	39\\
34	40\\
34	41\\
34	42\\
34	43\\
34	44\\
34	45\\
34	46\\
34	47\\
34	48\\
35	34\\
35	35\\
35	36\\
35	37\\
35	38\\
35	39\\
35	40\\
35	41\\
35	42\\
35	43\\
35	44\\
35	45\\
35	46\\
35	47\\
35	48\\
36	34\\
36	35\\
36	36\\
36	37\\
36	38\\
36	39\\
36	40\\
36	41\\
36	42\\
36	43\\
36	44\\
36	45\\
36	46\\
36	47\\
36	48\\
37	34\\
37	35\\
37	36\\
37	37\\
37	38\\
37	39\\
37	40\\
37	41\\
37	42\\
37	43\\
37	44\\
37	45\\
37	46\\
37	47\\
37	48\\
38	34\\
38	35\\
38	36\\
38	37\\
38	38\\
38	39\\
38	40\\
38	41\\
38	42\\
38	43\\
38	44\\
38	45\\
38	46\\
38	47\\
38	48\\
39	34\\
39	35\\
39	36\\
39	37\\
39	38\\
39	39\\
39	40\\
39	41\\
39	42\\
39	43\\
39	44\\
39	45\\
39	46\\
39	47\\
39	48\\
40	34\\
40	35\\
40	36\\
40	37\\
40	38\\
40	39\\
40	40\\
40	41\\
40	42\\
40	43\\
40	44\\
40	45\\
40	46\\
40	47\\
40	48\\
41	34\\
41	35\\
41	36\\
41	37\\
41	38\\
41	39\\
41	40\\
41	41\\
41	42\\
41	43\\
41	44\\
41	45\\
41	46\\
41	47\\
41	48\\
42	34\\
42	35\\
42	36\\
42	37\\
42	38\\
42	39\\
42	40\\
42	41\\
42	42\\
42	43\\
42	44\\
42	45\\
42	46\\
42	47\\
42	48\\
49	55\\
49	56\\
49	57\\
49	58\\
49	59\\
49	60\\
49	61\\
49	62\\
49	63\\
49	64\\
49	65\\
49	66\\
49	67\\
49	68\\
49	69\\
50	55\\
50	56\\
50	57\\
50	58\\
50	59\\
50	60\\
50	61\\
50	62\\
50	63\\
50	64\\
50	65\\
50	66\\
50	67\\
50	68\\
50	69\\
51	55\\
51	56\\
51	57\\
51	58\\
51	59\\
51	60\\
51	61\\
51	62\\
51	63\\
51	64\\
51	65\\
51	66\\
51	67\\
51	68\\
51	69\\
52	55\\
52	56\\
52	57\\
52	58\\
52	59\\
52	60\\
52	61\\
52	62\\
52	63\\
52	64\\
52	65\\
52	66\\
52	67\\
52	68\\
52	69\\
53	55\\
53	56\\
53	57\\
53	58\\
53	59\\
53	60\\
53	61\\
53	62\\
53	63\\
53	64\\
53	65\\
53	66\\
53	67\\
53	68\\
53	69\\
54	55\\
54	56\\
54	57\\
54	58\\
54	59\\
54	60\\
54	61\\
54	62\\
54	63\\
54	64\\
54	65\\
54	66\\
54	67\\
54	68\\
54	69\\
55	55\\
55	56\\
55	57\\
55	58\\
55	59\\
55	60\\
55	61\\
55	62\\
55	63\\
55	64\\
55	65\\
55	66\\
55	67\\
55	68\\
55	69\\
56	55\\
56	56\\
56	57\\
56	58\\
56	59\\
56	60\\
56	61\\
56	62\\
56	63\\
56	64\\
56	65\\
56	66\\
56	67\\
56	68\\
56	69\\
57	55\\
57	56\\
57	57\\
57	58\\
57	59\\
57	60\\
57	61\\
57	62\\
57	63\\
57	64\\
57	65\\
57	66\\
57	67\\
57	68\\
57	69\\
58	55\\
58	56\\
58	57\\
58	58\\
58	59\\
58	60\\
58	61\\
58	62\\
58	63\\
58	64\\
58	65\\
58	66\\
58	67\\
58	68\\
58	69\\
59	55\\
59	56\\
59	57\\
59	58\\
59	59\\
59	60\\
59	61\\
59	62\\
59	63\\
59	64\\
59	65\\
59	66\\
59	67\\
59	68\\
59	69\\
60	55\\
60	56\\
60	57\\
60	58\\
60	59\\
60	60\\
60	61\\
60	62\\
60	63\\
60	64\\
60	65\\
60	66\\
60	67\\
60	68\\
60	69\\
61	55\\
61	56\\
61	57\\
61	58\\
61	59\\
61	60\\
61	61\\
61	62\\
61	63\\
61	64\\
61	65\\
61	66\\
61	67\\
61	68\\
61	69\\
62	55\\
62	56\\
62	57\\
62	58\\
62	59\\
62	60\\
62	61\\
62	62\\
62	63\\
62	64\\
62	65\\
62	66\\
62	67\\
62	68\\
62	69\\
63	55\\
63	56\\
63	57\\
63	58\\
63	59\\
63	60\\
63	61\\
63	62\\
63	63\\
63	64\\
63	65\\
63	66\\
63	67\\
63	68\\
63	69\\
70	76\\
70	77\\
70	78\\
70	79\\
70	80\\
70	81\\
70	82\\
70	83\\
70	84\\
70	85\\
70	86\\
70	87\\
70	88\\
70	89\\
70	90\\
71	76\\
71	77\\
71	78\\
71	79\\
71	80\\
71	81\\
71	82\\
71	83\\
71	84\\
71	85\\
71	86\\
71	87\\
71	88\\
71	89\\
71	90\\
72	76\\
72	77\\
72	78\\
72	79\\
72	80\\
72	81\\
72	82\\
72	83\\
72	84\\
72	85\\
72	86\\
72	87\\
72	88\\
72	89\\
72	90\\
73	76\\
73	77\\
73	78\\
73	79\\
73	80\\
73	81\\
73	82\\
73	83\\
73	84\\
73	85\\
73	86\\
73	87\\
73	88\\
73	89\\
73	90\\
74	76\\
74	77\\
74	78\\
74	79\\
74	80\\
74	81\\
74	82\\
74	83\\
74	84\\
74	85\\
74	86\\
74	87\\
74	88\\
74	89\\
74	90\\
75	76\\
75	77\\
75	78\\
75	79\\
75	80\\
75	81\\
75	82\\
75	83\\
75	84\\
75	85\\
75	86\\
75	87\\
75	88\\
75	89\\
75	90\\
76	76\\
76	77\\
76	78\\
76	79\\
76	80\\
76	81\\
76	82\\
76	83\\
76	84\\
76	85\\
76	86\\
76	87\\
76	88\\
76	89\\
76	90\\
77	76\\
77	77\\
77	78\\
77	79\\
77	80\\
77	81\\
77	82\\
77	83\\
77	84\\
77	85\\
77	86\\
77	87\\
77	88\\
77	89\\
77	90\\
78	76\\
78	77\\
78	78\\
78	79\\
78	80\\
78	81\\
78	82\\
78	83\\
78	84\\
78	85\\
78	86\\
78	87\\
78	88\\
78	89\\
78	90\\
79	76\\
79	77\\
79	78\\
79	79\\
79	80\\
79	81\\
79	82\\
79	83\\
79	84\\
79	85\\
79	86\\
79	87\\
79	88\\
79	89\\
79	90\\
80	76\\
80	77\\
80	78\\
80	79\\
80	80\\
80	81\\
80	82\\
80	83\\
80	84\\
80	85\\
80	86\\
80	87\\
80	88\\
80	89\\
80	90\\
81	76\\
81	77\\
81	78\\
81	79\\
81	80\\
81	81\\
81	82\\
81	83\\
81	84\\
81	85\\
81	86\\
81	87\\
81	88\\
81	89\\
81	90\\
82	76\\
82	77\\
82	78\\
82	79\\
82	80\\
82	81\\
82	82\\
82	83\\
82	84\\
82	85\\
82	86\\
82	87\\
82	88\\
82	89\\
82	90\\
83	76\\
83	77\\
83	78\\
83	79\\
83	80\\
83	81\\
83	82\\
83	83\\
83	84\\
83	85\\
83	86\\
83	87\\
83	88\\
83	89\\
83	90\\
84	76\\
84	77\\
84	78\\
84	79\\
84	80\\
84	81\\
84	82\\
84	83\\
84	84\\
84	85\\
84	86\\
84	87\\
84	88\\
84	89\\
84	90\\
91	97\\
91	98\\
91	99\\
91	100\\
91	101\\
91	102\\
91	103\\
91	104\\
91	105\\
91	106\\
91	107\\
91	108\\
91	109\\
91	110\\
91	111\\
92	97\\
92	98\\
92	99\\
92	100\\
92	101\\
92	102\\
92	103\\
92	104\\
92	105\\
92	106\\
92	107\\
92	108\\
92	109\\
92	110\\
92	111\\
93	97\\
93	98\\
93	99\\
93	100\\
93	101\\
93	102\\
93	103\\
93	104\\
93	105\\
93	106\\
93	107\\
93	108\\
93	109\\
93	110\\
93	111\\
94	97\\
94	98\\
94	99\\
94	100\\
94	101\\
94	102\\
94	103\\
94	104\\
94	105\\
94	106\\
94	107\\
94	108\\
94	109\\
94	110\\
94	111\\
95	97\\
95	98\\
95	99\\
95	100\\
95	101\\
95	102\\
95	103\\
95	104\\
95	105\\
95	106\\
95	107\\
95	108\\
95	109\\
95	110\\
95	111\\
96	97\\
96	98\\
96	99\\
96	100\\
96	101\\
96	102\\
96	103\\
96	104\\
96	105\\
96	106\\
96	107\\
96	108\\
96	109\\
96	110\\
96	111\\
97	97\\
97	98\\
97	99\\
97	100\\
97	101\\
97	102\\
97	103\\
97	104\\
97	105\\
97	106\\
97	107\\
97	108\\
97	109\\
97	110\\
97	111\\
98	97\\
98	98\\
98	99\\
98	100\\
98	101\\
98	102\\
98	103\\
98	104\\
98	105\\
98	106\\
98	107\\
98	108\\
98	109\\
98	110\\
98	111\\
99	97\\
99	98\\
99	99\\
99	100\\
99	101\\
99	102\\
99	103\\
99	104\\
99	105\\
99	106\\
99	107\\
99	108\\
99	109\\
99	110\\
99	111\\
100	97\\
100	98\\
100	99\\
100	100\\
100	101\\
100	102\\
100	103\\
100	104\\
100	105\\
100	106\\
100	107\\
100	108\\
100	109\\
100	110\\
100	111\\
101	97\\
101	98\\
101	99\\
101	100\\
101	101\\
101	102\\
101	103\\
101	104\\
101	105\\
101	106\\
101	107\\
101	108\\
101	109\\
101	110\\
101	111\\
102	97\\
102	98\\
102	99\\
102	100\\
102	101\\
102	102\\
102	103\\
102	104\\
102	105\\
102	106\\
102	107\\
102	108\\
102	109\\
102	110\\
102	111\\
103	97\\
103	98\\
103	99\\
103	100\\
103	101\\
103	102\\
103	103\\
103	104\\
103	105\\
103	106\\
103	107\\
103	108\\
103	109\\
103	110\\
103	111\\
104	97\\
104	98\\
104	99\\
104	100\\
104	101\\
104	102\\
104	103\\
104	104\\
104	105\\
104	106\\
104	107\\
104	108\\
104	109\\
104	110\\
104	111\\
105	97\\
105	98\\
105	99\\
105	100\\
105	101\\
105	102\\
105	103\\
105	104\\
105	105\\
105	106\\
105	107\\
105	108\\
105	109\\
105	110\\
105	111\\
};

\addplot [color=Bittersweet, draw=none, mark size=0.5pt, mark=o, mark options={solid, fill=white, Bittersweet}]
  table[row sep=crcr]{%
22	7\\
23	8\\
24	9\\
25	10\\
26	11\\
27	12\\
43	28\\
44	29\\
45	30\\
46	31\\
47	32\\
48	33\\
64	49\\
65	50\\
66	51\\
67	52\\
68	53\\
69	54\\
85	70\\
86	71\\
87	72\\
88	73\\
89	74\\
90	75\\
106	91\\
107	92\\
108	93\\
109	94\\
110	95\\
111	96\\
};

\addplot [color=black, draw=none, mark size=0.5pt, mark=o, mark options={solid, black}]
  table[row sep=crcr]{%
1	1\\
1	13\\
1	14\\
1	15\\
1	19\\
1	23\\
2	2\\
2	13\\
2	15\\
2	16\\
2	19\\
2	23\\
3	3\\
3	15\\
3	17\\
3	18\\
3	19\\
3	23\\
4	4\\
4	15\\
4	17\\
4	19\\
4	20\\
4	23\\
5	5\\
5	15\\
5	19\\
5	21\\
5	22\\
5	23\\
6	6\\
6	15\\
6	19\\
6	21\\
6	23\\
6	24\\
7	13\\
7	14\\
7	16\\
7	20\\
7	24\\
8	14\\
8	15\\
8	16\\
8	20\\
8	24\\
9	16\\
9	17\\
9	18\\
9	20\\
9	24\\
10	16\\
10	18\\
10	19\\
10	20\\
10	24\\
11	16\\
11	20\\
11	21\\
11	22\\
11	24\\
12	16\\
12	20\\
12	22\\
12	23\\
12	24\\
13	15\\
13	16\\
13	19\\
13	20\\
13	23\\
13	24\\
13	25\\
14	15\\
14	16\\
14	19\\
14	20\\
14	23\\
14	24\\
14	26\\
15	15\\
15	16\\
15	19\\
15	20\\
15	23\\
15	24\\
15	27\\
16	7\\
16	13\\
16	14\\
17	8\\
17	15\\
17	16\\
18	9\\
18	17\\
18	18\\
19	10\\
19	19\\
19	20\\
20	11\\
20	21\\
20	22\\
21	12\\
21	23\\
21	24\\
22	7\\
22	34\\
22	35\\
22	36\\
22	40\\
22	44\\
23	8\\
23	34\\
23	36\\
23	37\\
23	40\\
23	44\\
24	9\\
24	36\\
24	38\\
24	39\\
24	40\\
24	44\\
25	10\\
25	36\\
25	38\\
25	40\\
25	41\\
25	44\\
26	11\\
26	36\\
26	40\\
26	42\\
26	43\\
26	44\\
27	12\\
27	36\\
27	40\\
27	42\\
27	44\\
27	45\\
28	34\\
28	35\\
28	37\\
28	41\\
28	45\\
29	35\\
29	36\\
29	37\\
29	41\\
29	45\\
30	37\\
30	38\\
30	39\\
30	41\\
30	45\\
31	37\\
31	39\\
31	40\\
31	41\\
31	45\\
32	37\\
32	41\\
32	42\\
32	43\\
32	45\\
33	37\\
33	41\\
33	43\\
33	44\\
33	45\\
34	36\\
34	37\\
34	40\\
34	41\\
34	44\\
34	45\\
34	46\\
35	36\\
35	37\\
35	40\\
35	41\\
35	44\\
35	45\\
35	47\\
36	36\\
36	37\\
36	40\\
36	41\\
36	44\\
36	45\\
36	48\\
37	28\\
37	34\\
37	35\\
38	29\\
38	36\\
38	37\\
39	30\\
39	38\\
39	39\\
40	31\\
40	40\\
40	41\\
41	32\\
41	42\\
41	43\\
42	33\\
42	44\\
42	45\\
43	28\\
43	55\\
43	56\\
43	57\\
43	61\\
43	65\\
44	29\\
44	55\\
44	57\\
44	58\\
44	61\\
44	65\\
45	30\\
45	57\\
45	59\\
45	60\\
45	61\\
45	65\\
46	31\\
46	57\\
46	59\\
46	61\\
46	62\\
46	65\\
47	32\\
47	57\\
47	61\\
47	63\\
47	64\\
47	65\\
48	33\\
48	57\\
48	61\\
48	63\\
48	65\\
48	66\\
49	55\\
49	56\\
49	58\\
49	62\\
49	66\\
50	56\\
50	57\\
50	58\\
50	62\\
50	66\\
51	58\\
51	59\\
51	60\\
51	62\\
51	66\\
52	58\\
52	60\\
52	61\\
52	62\\
52	66\\
53	58\\
53	62\\
53	63\\
53	64\\
53	66\\
54	58\\
54	62\\
54	64\\
54	65\\
54	66\\
55	57\\
55	58\\
55	61\\
55	62\\
55	65\\
55	66\\
55	67\\
56	57\\
56	58\\
56	61\\
56	62\\
56	65\\
56	66\\
56	68\\
57	57\\
57	58\\
57	61\\
57	62\\
57	65\\
57	66\\
57	69\\
58	49\\
58	55\\
58	56\\
59	50\\
59	57\\
59	58\\
60	51\\
60	59\\
60	60\\
61	52\\
61	61\\
61	62\\
62	53\\
62	63\\
62	64\\
63	54\\
63	65\\
63	66\\
64	49\\
64	76\\
64	77\\
64	78\\
64	82\\
64	86\\
65	50\\
65	76\\
65	78\\
65	79\\
65	82\\
65	86\\
66	51\\
66	78\\
66	80\\
66	81\\
66	82\\
66	86\\
67	52\\
67	78\\
67	80\\
67	82\\
67	83\\
67	86\\
68	53\\
68	78\\
68	82\\
68	84\\
68	85\\
68	86\\
69	54\\
69	78\\
69	82\\
69	84\\
69	86\\
69	87\\
70	76\\
70	77\\
70	79\\
70	83\\
70	87\\
71	77\\
71	78\\
71	79\\
71	83\\
71	87\\
72	79\\
72	80\\
72	81\\
72	83\\
72	87\\
73	79\\
73	81\\
73	82\\
73	83\\
73	87\\
74	79\\
74	83\\
74	84\\
74	85\\
74	87\\
75	79\\
75	83\\
75	85\\
75	86\\
75	87\\
76	78\\
76	79\\
76	82\\
76	83\\
76	86\\
76	87\\
76	88\\
77	78\\
77	79\\
77	82\\
77	83\\
77	86\\
77	87\\
77	89\\
78	78\\
78	79\\
78	82\\
78	83\\
78	86\\
78	87\\
78	90\\
79	70\\
79	76\\
79	77\\
80	71\\
80	78\\
80	79\\
81	72\\
81	80\\
81	81\\
82	73\\
82	82\\
82	83\\
83	74\\
83	84\\
83	85\\
84	75\\
84	86\\
84	87\\
85	70\\
85	97\\
85	98\\
85	99\\
85	103\\
85	107\\
86	71\\
86	97\\
86	99\\
86	100\\
86	103\\
86	107\\
87	72\\
87	99\\
87	101\\
87	102\\
87	103\\
87	107\\
88	73\\
88	99\\
88	101\\
88	103\\
88	104\\
88	107\\
89	74\\
89	99\\
89	103\\
89	105\\
89	106\\
89	107\\
90	75\\
90	99\\
90	103\\
90	105\\
90	107\\
90	108\\
91	97\\
91	98\\
91	100\\
91	104\\
91	108\\
92	98\\
92	99\\
92	100\\
92	104\\
92	108\\
93	100\\
93	101\\
93	102\\
93	104\\
93	108\\
94	100\\
94	102\\
94	103\\
94	104\\
94	108\\
95	100\\
95	104\\
95	105\\
95	106\\
95	108\\
96	100\\
96	104\\
96	106\\
96	107\\
96	108\\
97	99\\
97	100\\
97	103\\
97	104\\
97	107\\
97	108\\
97	109\\
98	99\\
98	100\\
98	103\\
98	104\\
98	107\\
98	108\\
98	110\\
99	99\\
99	100\\
99	103\\
99	104\\
99	107\\
99	108\\
99	111\\
100	91\\
100	97\\
100	98\\
101	92\\
101	99\\
101	100\\
102	93\\
102	101\\
102	102\\
103	94\\
103	103\\
103	104\\
104	95\\
104	105\\
104	106\\
105	96\\
105	107\\
105	108\\
106	91\\
107	92\\
108	93\\
109	94\\
110	95\\
111	96\\
106	112\\
107	112\\
108	112\\
109	112\\
110	112\\
111	112\\
};

  \addplot [color=BurntOrange, draw=none, mark size=0.25pt, mark=*, mark options={solid, fill=white, BurntOrange}]
  table[row sep=crcr]{%
  1	1\\
  2	2\\
  3	3\\
  4	4\\
  5	5\\
  6	6\\
  106	112\\
  107	112\\
  108	112\\
  109	112\\
  110	112\\
  111	112\\
  };
\end{axis}
\end{tikzpicture}%

%% file: figures/t_sampling/t_sampling.tex
\begin{tikzpicture}


\def\T{5}
\def\dtMPC{0.1}

\def\NintA{4}
\def\NintB{8}
\def\NintC{10}

\begin{axis}[
    width=0.92\linewidth,
    height=4.5cm,
    xmin=0,
    xmax=\T,
    ymin=-0.5,
    ymax=2.5,
    axis lines=none,
    xtick={0,2,4,6,8,10},
    xticklabels={$0$,$2$,$4$,$6$,$8$,$10$},
    ytick=\empty,
    xlabel={$t$},
    xlabel style={
        at={(axis description cs:1.02,0)},
        anchor=west
    },
    clip=false,
]


\addplot[
    black,
    thick,
] coordinates {(0,2) (\T,2)};

\foreach \l in {1,...,3} {

    \pgfmathsetmacro{\texp}{
        floor(
            (\T/\dtMPC + 1)^(
                (\l-1)/\NintA
            )
        ) * \dtMPC
    }

    \pgfmathsetmacro{\tlin}{
        \l * \dtMPC
    }

    \pgfmathparse{
        \tlin > \texp ? \tlin : \texp
    }
    \let\tpoint\pgfmathresult

    \addplot[
        only marks,
        mark=*,
        mark size=1.5pt,
    ] coordinates {(\tpoint,2)};
}

\addplot[
    only marks,
    mark=*,
    mark size=1.5pt,
] coordinates {(\T,2)};

\node[anchor=east] at (axis cs:-0.15,2)
    {$N_{\mathrm{int}}=4$};


\addplot[
    black,
    thick,
] coordinates {(0,1) (\T,1)};

\foreach \l in {1,...,7} {

    \pgfmathsetmacro{\texp}{
        floor(
            (\T/\dtMPC + 1)^(
                (\l-1)/\NintB
            )
        ) * \dtMPC
    }

    \pgfmathsetmacro{\tlin}{
        \l * \dtMPC
    }

    \pgfmathparse{
        \tlin > \texp ? \tlin : \texp
    }
    \let\tpoint\pgfmathresult

    \addplot[
        only marks,
        mark=*,
        mark size=1.5pt,
    ] coordinates {(\tpoint,1)};
}

\addplot[
    only marks,
    mark=*,
    mark size=1.5pt,
] coordinates {(\T,1)};

\node[anchor=east] at (axis cs:-0.15,1)
    {$N_{\mathrm{int}}=8$};


\addplot[
    black,
    thick,
] coordinates {(0,0) (\T,0)};

\foreach \l in {1,...,9} {

    \pgfmathsetmacro{\texp}{
        floor(
            (\T/\dtMPC + 1)^(
                (\l-1)/\NintC
            )
        ) * \dtMPC
    }

    \pgfmathsetmacro{\tlin}{
        \l * \dtMPC
    }

    \pgfmathparse{
        \tlin > \texp ? \tlin : \texp
    }
    \let\tpoint\pgfmathresult

    \addplot[
        only marks,
        mark=*,
        mark size=1.5pt,
    ] coordinates {(\tpoint,0)};
}

\addplot[
    only marks,
    mark=*,
    mark size=1.5pt,
] coordinates {(\T,0)};

\node[anchor=east] at (axis cs:-0.15,0)
    {$N_{\mathrm{int}}=10$};


\node[anchor=center] at (axis cs:0,-0.5)
    {$0$};

\node[anchor=center] at (axis cs:\T,-0.5)
    {$T$};

\draw[thick]
    (axis cs:0,-0.2) -- (axis cs:0,2.2);

\draw[thick]
    (axis cs:\T,-0.2) -- (axis cs:\T,2.2);

\end{axis}

\end{tikzpicture}